\documentclass[a4paper,usenatbib]{mnras}

\usepackage{newtxtext}

\usepackage[T1]{fontenc}
\usepackage{ae,aecompl}

\usepackage{amsmath,amstext,amsgen,amsbsy,amsopn,amsfonts,theorem}

\usepackage[pdftex]{graphicx}
\usepackage{xspace}
\usepackage{amsmath}
\usepackage{hyperref}
\usepackage{framed}
\usepackage{txfonts}
\usepackage{epstopdf}
\usepackage{gensymb}

\usepackage[normalem]{ulem}

\usepackage{graphicx}	
\usepackage{amsmath}	
\usepackage{amssymb}	

\title[A fork in the Sagittarius trailing debris]{A fork in the
  Sagittarius trailing debris}

\author[C. Navarrete et al.]{C. Navarrete$^{1,2,3}$\thanks{Contact e-mail: 
\href{cnavarre@astro.puc.cl}{cnavarre@astro.puc.cl}}, V. Belokurov$^{3}$, S. E. 
Koposov$^{3}$, M. Irwin$^{3}$, M. Catelan$^{1,2}$, \newauthor S. 
Duffau$^{2,1}$ and A. J. Drake$^{4}$\\
$^{1}$Instituto de Astrof\'isica, Pontificia Universidad Cat\'olica de Chile, 
Av. Vicu\~na Mackenna 4860, 782-0436 Macul, Santiago, Chile \\ 
$^{2}$Millennium Institute of Astrophysics, Av. Vicu\~na Mackenna 4860, 
782-0436, Santiago, Chile \\
$^{3}$Institute of Astronomy, University of Cambridge, Madingley Road, 
Cambridge, CB3 0HA, UK\\
$^{4}$Cahill Center for Astronomy and Astrophysics, California Institute of Technology, Pasadena, CA 91125}

\date{Last updated 2016 xxxx xx; in original form 2016 xxxx xx}

\pubyear{2016}

\begin{document}
\label{firstpage}
\pagerange{\pageref{firstpage}--\pageref{lastpage}}
\maketitle

\begin{abstract}

We take advantage of the deep and wide coverage of the VST ATLAS
survey to study the line-of-sight structure of the Sagittarius
stellar stream in the Southern hemisphere, only $\sim40^{\circ}$ away
from the progenitor. We use photometrically selected Sub-Giant Branch
(SGB) stars to reveal a complex debris morphology of the trailing arm and
detect at least two clear peaks in the SGB distance modulus
distribution. The separation between the two line-of-sight components
is at least 5 kpc at the edge of the VST ATLAS footprint, but appears
to change along the stream, which allows us to conclude that these
detections correspond to two physically independent stellar
structures, rather than a mix of co-distant stellar populations within
a single stream. Our discovery of a fork in the Sgr trailing arm is
verified using Blue Horizontal Branch stars and our distance
measurements are calibrated using RR Lyrae stars from the Catalina
Real-Time Transient Survey. Comparing with numerical simulations
of the Sgr dwarf disruption, the more distant of the two components in the 
fork matches perfectly with the track of the trailing debris. However, no 
obvious counterpart exists in the
simulation for the closer line-of-sight component.

\end{abstract}




\section{Introduction}
More than a decade has passed since the vast expanse of the
Sagittarius (Sgr) tidal stream was uncovered by \citet{M03} through
dexterous data-mining of the 2MASS stellar catalogs. While the
M-giants used in that study are perfectly suitable as a tracer to map out
the extent of the stream, the full complexity of this structure
remained under-appreciated until more numerous and less
metallicity-biased tracers were deployed. Main Sequence (MS) and Main
Sequence Turn-Off (MSTO) stars outnumber M-giants by several orders of
magnitude and thus can help to uncover lower level sub-structures
within the tails. To that end, using the Sloan Digital Sky Survey
(SDSS) Data Release 5 (DR5) multi-band photometry, \citet{B06} showed
that around the North Galactic cap the Sgr leading tail is bifurcated,
i.e. split into two distinct components on the sky, a bright and a
faint one, running alongside each other in distance. As the SDSS
progressed from DR5 to DR9, a more complete view of the stream was put
together by \citet[][]{K12}, unveiling a similar bifurcation in the
trailing tail in the Southern Galactic hemisphere. This was later
confirmed by \citet{S13} in an independent study using data from
the Pan-STARRS1 survey. We note that both \citet{K12} and \citet{S13}
successfully used Red Clump (RC) stars as distance indicators
in their studies of the stream.

\begin{figure*}
  \includegraphics[width=\textwidth]{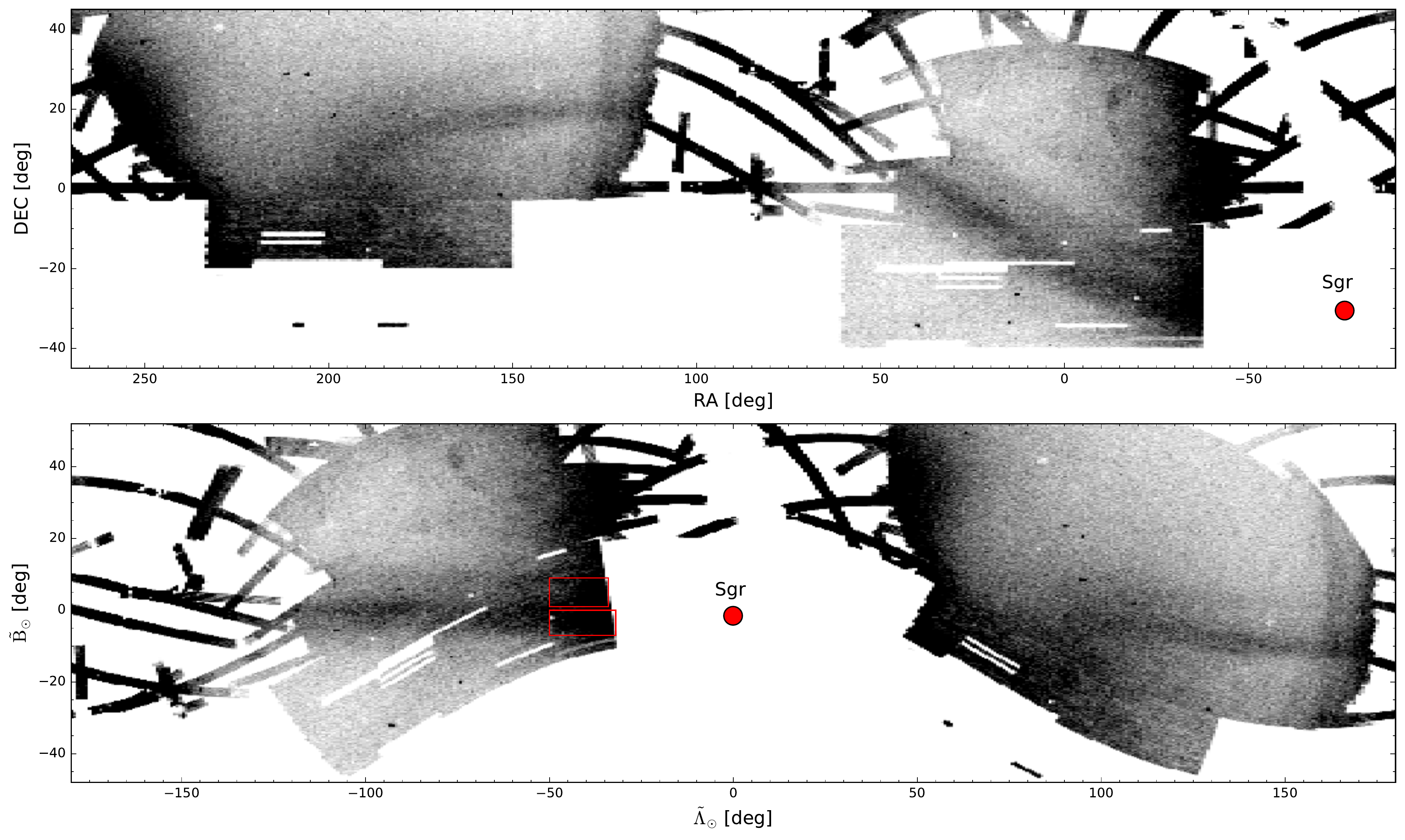}
  \caption{Density map of MSTO stars from the SDSS DR9 and VST ATLAS
    datasets, in equatorial (top) and the Sgr stream (bottom)
    coordinate systems. The darker regions correspond to enhanced
    stellar density. The red boxes enclose the area used to study the
    line of sight sub-structure on the Southern extension of the Sgr
    stream, for the bright (at $\tilde{\rm B}_{\odot} < 0^{\circ}$)
    and faint ($\tilde{\rm B}_{\odot} > 0^{\circ}$) branches.}
    \label{fig:stitch}
\end{figure*}

The final SDSS footprint has left two large gaps in the Sgr stream
coverage: one, at least $100^{\circ}$ wide in the immediate
neighborhood of the progenitor, and one of a similar size in the
direction towards the Galactic anti-center where the stream crosses
the disk. Therefore, in these two regions, studies of the stream have
so far relied on 2MASS M-giants. Unfortunately these
comparatively metal-rich and relatively young stars may have been fogging the
view of the Sgr tails. For example, the second, fainter component of
the Sgr bifurcation does not appear to be traced by M-giants at all,
which could be due to the metallicity differences between the two
branches as explained in \citet{K12}. Additionally, M-giants are not
very useful distance indicators as their intrinsic luminosity
strongly depends on both age and metallicity. Instead, as shown by
\citet{B06}, Sub-Giant Branch (SGB) stars can be selected to provide
accurate relative distances, which can be then calibrated using less
numerous but more trustworthy Blue Horizontal Branch (BHB) stars
\citep[see e.g.][]{Yanny2009,Ruhland2011,B14} or RR Lyrae \citep[see
  e.g.][]{DrakeStream}.

As the Sun is located very close to the plane of the Sgr stream, the
information on the shape of the tidal tails and thus the orbit of the
progenitor is encoded in the run of the line-of-sight distances along
the stream. Therefore, high distance accuracy is required to
facilitate an unbiased gravitational potential inference. However,
results so far have lead to some curious conclusions. This is illustrated in
\citet{LM10} who find that only a triaxial Dark Matter (DM) halo
resembling a hockey puck and oriented perpendicular to the Galactic
disk can explain the distances along the Sgr leading arm. Another
example is the study of \citet{Gibbons14} who show that a very light
Milky Way is needed to reproduce the Sgr stream orbital precession
data obtained by \citet{B14}. Moreover, good distance precision is
necessary to unpick multiple wraps of the stream in a given
direction. For example, \citet{B06} detected multiple debris
components along the sight-lines towards the leading tail using SGB
stars as tracers. It is likely that the so-called C branch of the
stream they see behind the leading debris is in fact a part of the
trailing tail. \citet{Correnti2010} used RC stars to produce
further evidence for multiple wraps of the stream.

\begin{figure*}
  \includegraphics[width=\textwidth]{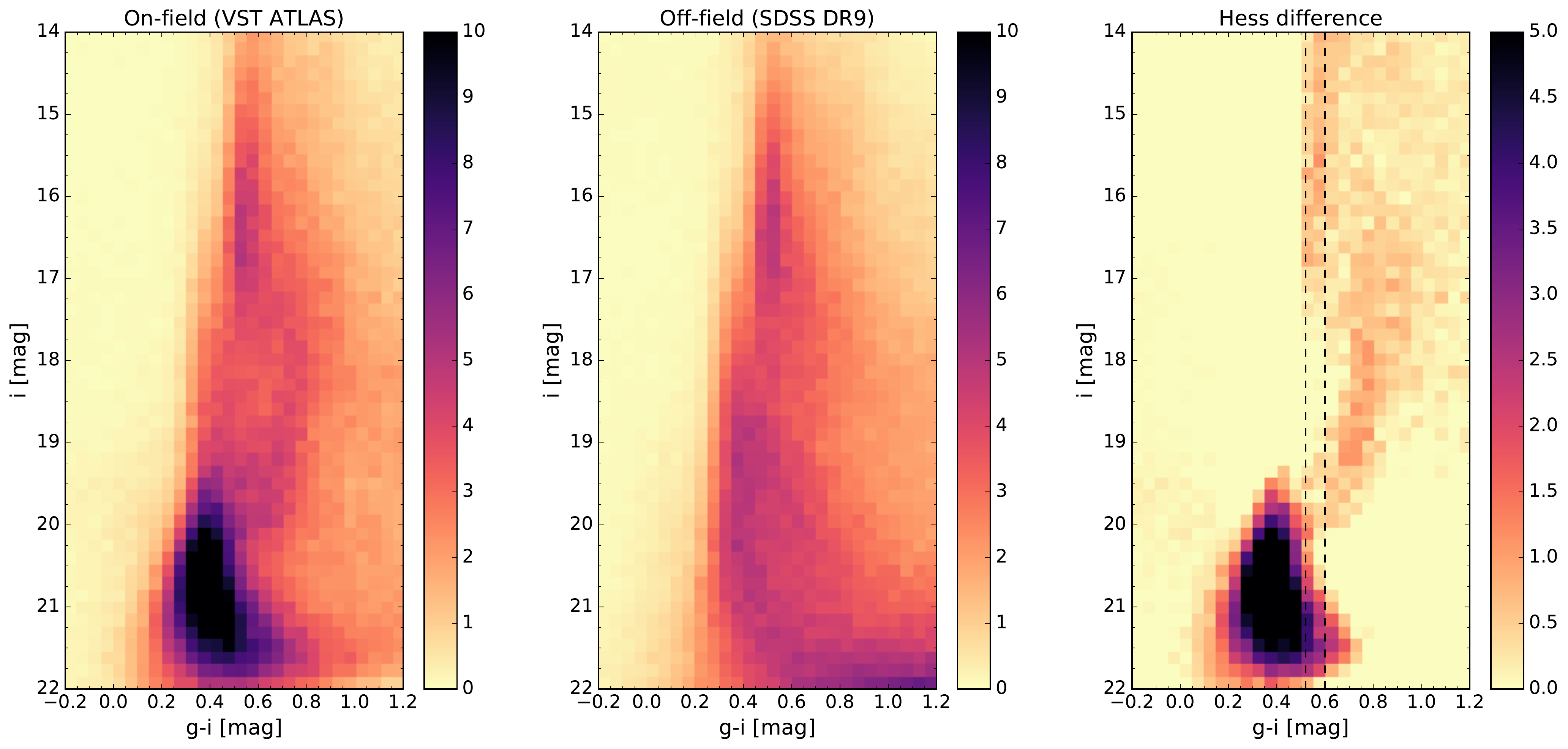}
  \caption{{\it Left:} Hess diagram for the on-stream field defined by
    --50$^{\circ}$ < $\tilde{\Lambda}_{\odot}$ < --32$^{\circ}$
    and --7$^{\circ} < \tilde{\rm B}_{\odot} < 0^{\circ}$. The MS and
    the MSTO of the Sgr population are both clearly visible at $i
    \sim$ 20 mag and fainter.  {\it Middle:} Hess diagram for the
    off-stream field, from SDSS data. {\it Right:}
    Background-subtracted Hess diagram of the Sgr stream. The MS,
    subgiant and red giant branches are all clearly recovered. The bin
    size in the Hess diagrams are 0.15 mag in $i$ magnitude and
    0.05 in $(g-i)$ color. The color bar scale corresponds to the
    number of stars per bin per square degree in each panel.}
    \label{fig:hess}
\end{figure*}

While difficult to track down, stream wraps offer powerful leverage in
tidal tail modelling. To begin with, they allow the minimal bound
on the stream length to be improved. As shown in \cite{Erkal16b},
the stream length is a function of the progenitor mass, the time of
disruption and the properties of the host gravitational
potential. Thus, if the progenitor original mass is constrained
independently \citep[see e.g.][]{MNO2010} and the time since the
beginning of disruption is deduced based on e.g. star-formation
activity in the stream \citep[][]{deBoer2015}, then the radial profile
of the Galactic total density can be inferred. Typically, the fainter
wraps of the stream are also the oldest detectable tidal
debris. These, therefore, have accumulated the largest amount of 
differential orbital plane precession. According to \cite{Erkal16a},
this is manifested in both the width of the stream on the sky as well
as the angular offset of the ``wrapped'' portion of the stream with
respect to the dynamically younger debris. The plane precession is
caused by the torques provided by the non-spherical portion of the
Galactic force field. Therefore, through the analysis of the older
stream wraps, a picture of the 3D shape of the Milky Way DM 
halo can be reconstructed.

\begin{figure*}
  \includegraphics[width=\textwidth]{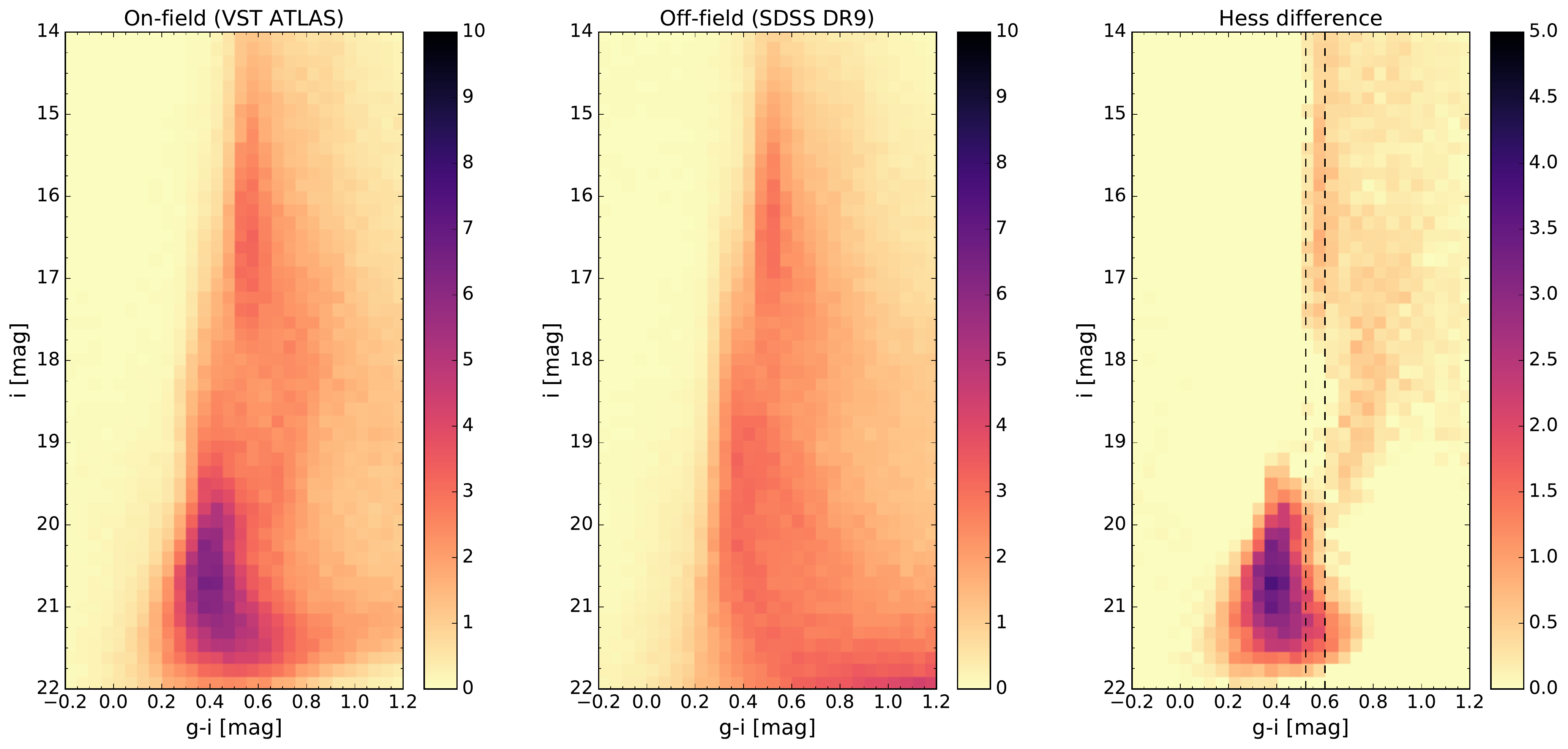}
  \caption{Same as Figure~\ref{fig:hess}, but for the region enclosed
    by --50$^{\circ}$ < $\tilde{\Lambda}_{\odot}$ < --34$^{\circ}$
    and 1$^{\circ} < \tilde{\rm B}_{\odot} < 9^{\circ}$,
    i.e. corresponding to the location of the faint branch of the Sgr
    trailing arm. The Hess difference hints at the existence of two
    SGBs, as in the bright branch, but with much less signal (compare with the 
    right panel of Figure~\ref{fig:hess}). }
    \label{fig:hess_faint}
\end{figure*}

\begin{figure*}
  \centering
  \includegraphics[width=0.45\textwidth]{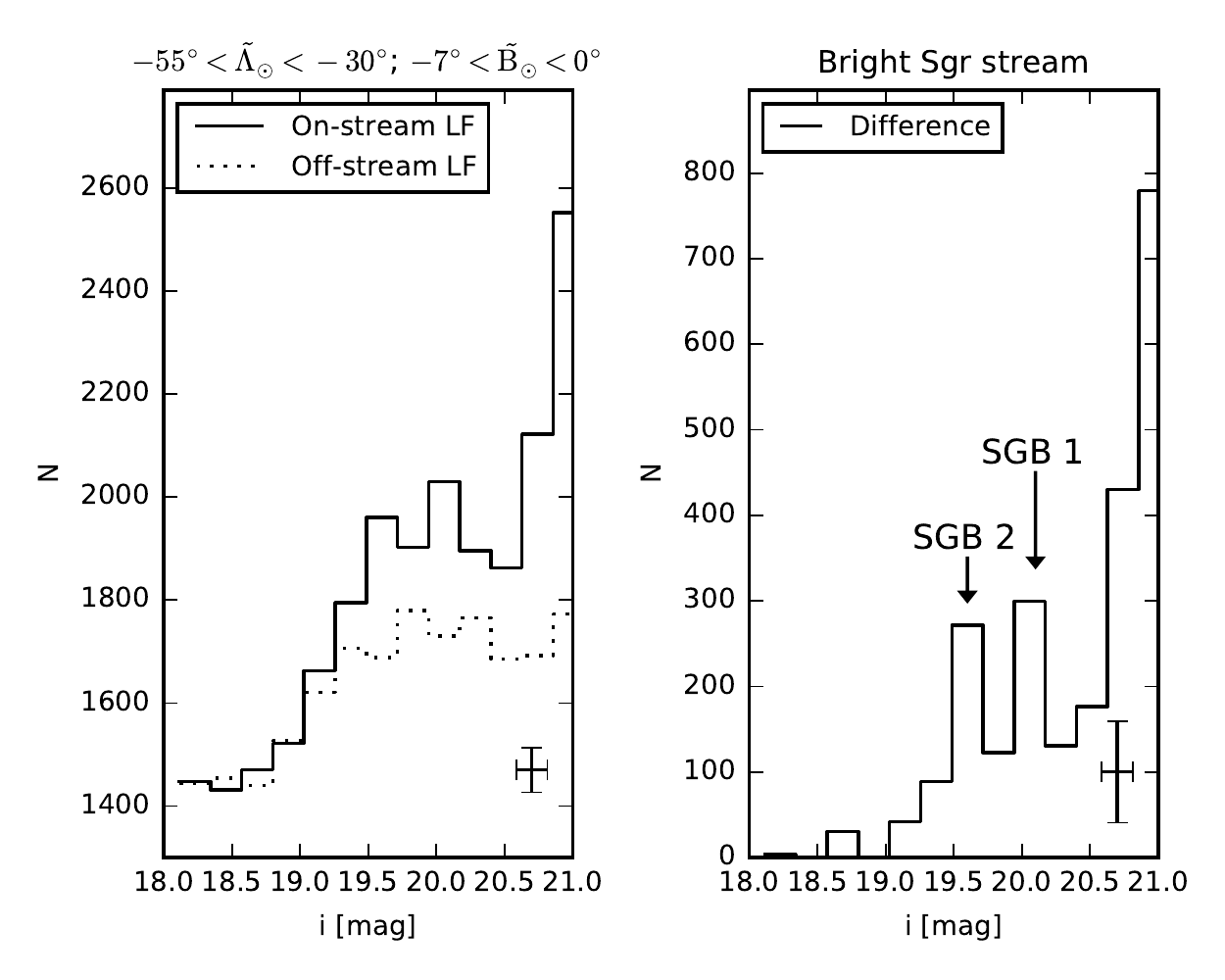}
  \includegraphics[width=0.45\textwidth]{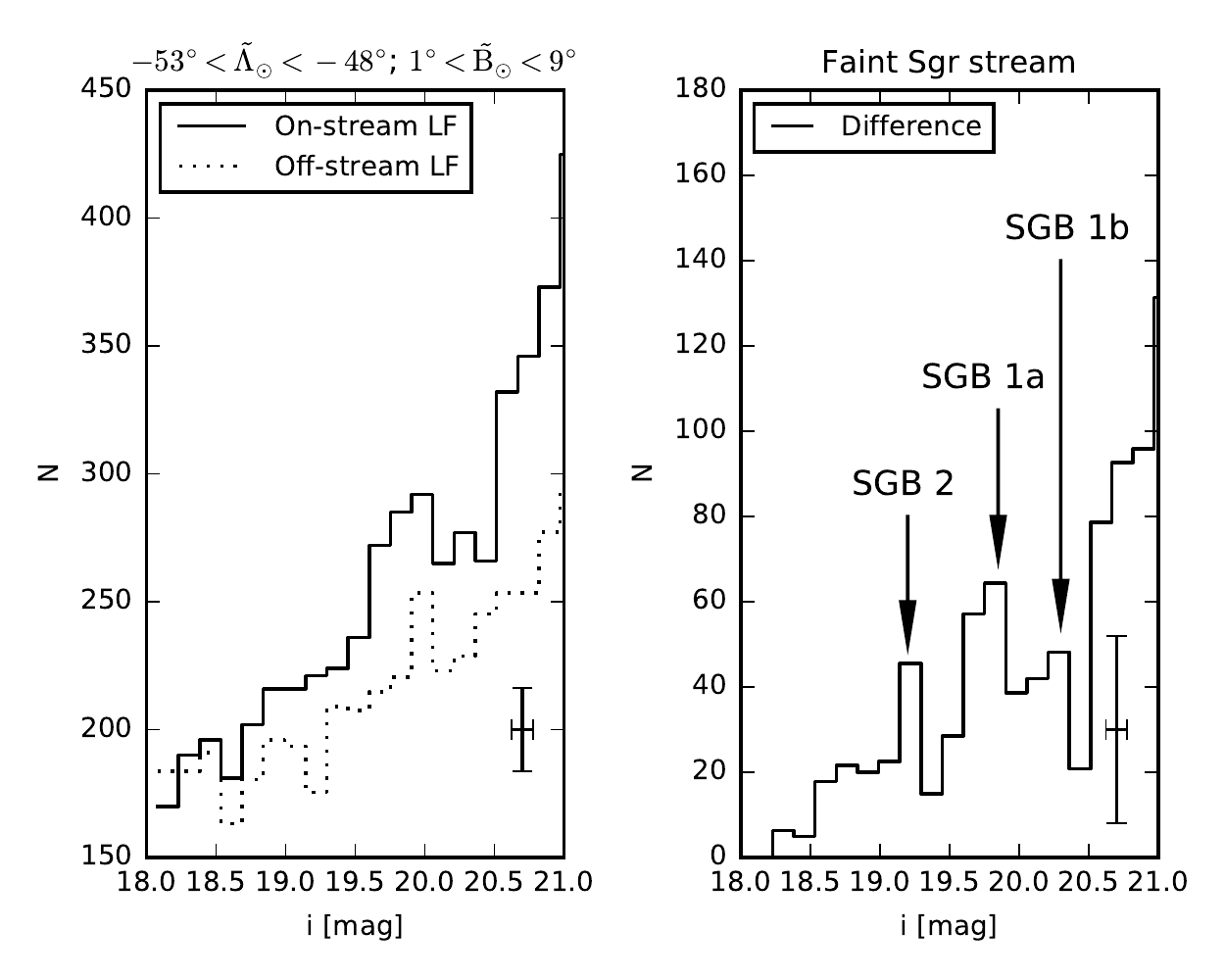}
   \caption{Luminosity functions of SGB stars selected to lie in a narrow color 
range in the bright branch (dashed lines in Figures~\ref{fig:hess} and 
\ref{fig:hess_faint}). {\it Left}: Luminosity functions for the SGB stars for 
the on-stream field (solid line) and the off-stream field (dotted line), scaled 
by the ratio of their areas for the bright Sgr arm, at $-55^{\circ} < 
\tilde{\Lambda}_{\odot} < -30^{\circ}$ and $-7^{\circ} < \tilde{\rm B}_{\odot} < 
0^{\circ}$. {\it Left middle:} Difference of the luminosity functions for the 
on- and the off-stream fields shown in the left panel. Note the two peaks 
visible at $i \sim$ 19.6 and 20.0 mag. {\it Right middle:} The same as leftmost 
panel, but for the area enclosing the Sgr faint branch: $-53^{\circ} < 
\tilde{\Lambda}_{\odot} < -48^{\circ}$ and $1^{\circ} < \tilde{\rm B}_{\odot} < 
9^{\circ}$. The distributions corresponds to the on-stream (solid line) and 
off-stream (dotted line) fields. {\it Right:} Difference of the luminosity 
functions for the on-stream region of the Sgr faint branch. Here three peaks are 
noticeable, at $i \sim 19.3$, $19.9$ and $20.3$ mag. Representative error bars 
are shown at the bottom right in each panel.}
    \label{fig:lf}
\end{figure*}

In this paper, we analyze the most recent imaging data from the VST
ATLAS survey, which covers a portion of the Sgr trailing tail in the Southern
hemisphere not previously observed by any wide-field optical
survey. In particular, we focus on the 3D behaviour of this un-studied
part of the stream and show that i) the trailing tail bifurcation
detected by \citet{K12} continues into the VST ATLAS footprint and ii)
the trailing debris appears to be split along the line-of-sight, with
the additional stream component following a distinct distance
track. This paper is structured as follows. Section~\ref{sec:vst_sgr}
describes the VST ATLAS data and presents Color-Magnitude Diagrams
(CMD) of the Sgr trailing tail. Section~\ref{sec:tomography} presents
the analysis of the CMD and the resulting relative distance
measurements. The absolute distance scale is introduced in
Section~\ref{sec:dist}. Discussion and Conclusions can be found in
Section~\ref{sec:dandc}.

\section{The Sgr stream in the ATLAS survey}
\label{sec:vst_sgr}

The ATLAS survey \citep{S15} is being carried out with the 2.6 m VLT Survey 
Telescope (VST) at ESO's Cerro Paranal Observatory in Chile. The survey uses the 
OmegaCAM camera, containing 32 CCDs of 4k $\times$ 2k pixels with a sampling
of 0.21$\arcsec$ per pixel giving a field-of-view coverage of 1 deg$^{2}$.
The ATLAS survey aims to cover 4700 deg$^2$ 
of the southern sky in five photometric passbands, $ugriz$, reaching depths 
similar to the SDSS (e.g., the 5$\sigma$ source detection limit is 23.1 mag for 
the $g$ band). Basic image reduction and initial catalogue generation are 
provided by the Cambridge Astronomical Survey Unit (CASU) using the VST data 
flow software \citep[for details see][]{S15}. The band-merging and selection of 
primary sources were performed as separate steps using a local SQL database 
\citep[as described in detail in][]{K14}. Note that the VST ATLAS photometry 
used in this work has been calibrated with the data from AAVSO Photometric 
All-Sky Survey \citep[APASS,][]{Henden2016}. While approximately similar to the 
SDSS, in practice the APASS $gri$ filters do not match the SDSS ones exactly. 
Moreover, APASS does not provide any $u$-band data, therefore our calibration 
of the VST ATLAS $u$-band magnitudes relies on the $g$ and $r$ photometry. 
In this work, all magnitudes are dereddened using the \cite{SFD98} extinction 
maps, adopting the extinction coefficients of \cite{SF11}.

Figure~\ref{fig:stitch} shows the spatial density distribution of MSTO stars in 
the SDSS DR9 \citep{A12} and the VST ATLAS surveys stitched together, both in 
equatorial (top panel) and Sagittarius-stream (bottom panel) 
coordinates\footnote{($\tilde{\Lambda}_{\odot}$, $\tilde{\rm B}_{\odot}$) are 
the coordinates of the Sgr orbital  plane, as defined by \cite{B14}.}. The MSTO 
stars were selected to have 0.0 $< (g-i) <$ 0.6 and 19 $< r <$ 21 mag in both 
datasets. The southern fingers of SDSS DR9, which overlap the observations of 
VST ATLAS, were excluded to not artificially increase the number of stars in 
those regions. VST ATLAS covers two big patches of the sky at negative 
declinations, centered at $(\alpha, \delta)$= (195.0$^{\circ}$, 
--24.5$^{\circ}$) ; and (10.0$^{\circ}$,--11.0$^{\circ}$). 

In the SDSS data, the two branches of the leading tail are visible crossing most 
of the SDSS footprint. The ATLAS survey observes the region of the Southern sky 
closer to the Sgr dwarf, thus extending the view of the trailing arm beyond what 
is available in the SDSS DR9 data. Note that \cite{K12} and \cite{S13} had 
already identified two branches in the trailing tail in the South, similar to 
what is seen in the leading tail in the North. The VST ATLAS data unambiguously 
confirms the existence of the trailing bifurcation at lower 
$|\tilde{\Lambda}_{\odot}|$ as evidenced by the bottom panel of the Figure. To 
recap, the bifurcation consists of two branches that are separated on the sky by 
$\sim$10$^{\circ}$. Note that, according to \citet[][]{S13}, there could also be 
a difference in the heliocentric distances to the two branches, of order 
$\sim$~5~kpc. The bulk of the ``bright stream'' is below the Sgr orbital plane 
(i.e., at $\tilde{\rm B}_{\odot} < 0^{\circ}$) while the ``faint stream'' lies 
mostly above the plane ($\tilde{\rm B}_{\odot} > 0^{\circ}$). While providing 
some additional information on the density evolution of the faint branch, the 
VST ATLAS covers mostly the region with $\tilde{\rm B}_{\odot} < 0^{\circ}$. 
Thus the focus of this study is mostly on the bright branch of the trailing Sgr 
tail.

\subsection{Color-magnitude diagrams}\label{cmd}

\begin{figure*}
  \centering
  \includegraphics[width=\textwidth]{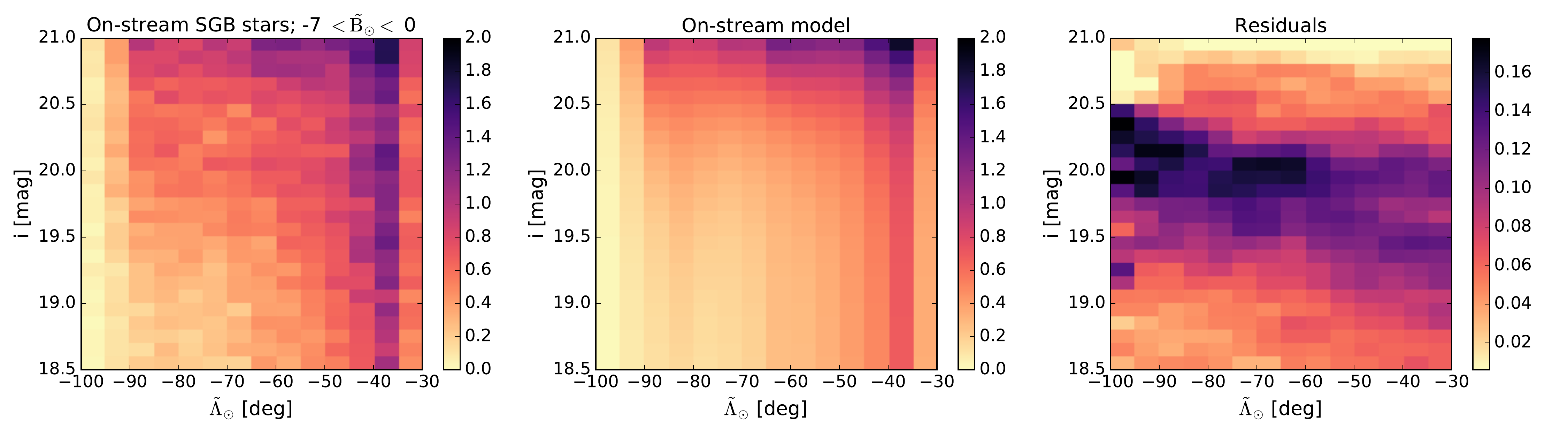}
  \caption{Density of the bright branch SGB stars as a function of the longitude 
along the stream, $\tilde{\Lambda}_{\odot}$, and the $i$-band magnitude. The 
on-stream field (left), the model (center) and the residuals (right) are shown. 
The 2D density was divided by the area covered by the field. The diagrams are 14 
$\times$ 25 pixels, and the residuals are smoothed with a Gaussian kernel with a 
FWHM of 0.7 pixels. At each $\tilde{\Lambda}_{\odot}$ bin, the residuals are 
vertically normalized. The two peaks in the Sgr bright stream (the ``fork'') 
start at $\tilde{\Lambda}_{\odot} \sim$--30$^{\circ}$, separated by $\sim$0.6 
mag, and tend to merge at $\tilde{\Lambda}_{\odot} \sim$--70$^{\circ}$. 
Afterwards, there are also two peaks, with the most prominent one located at 
i$\sim$20.2 mag, and another much less prominent at $\sim$19.4 mag.}
    \label{trace_sgb}
\end{figure*}
\begin{figure*}
  \centering
  \includegraphics[width=\textwidth]{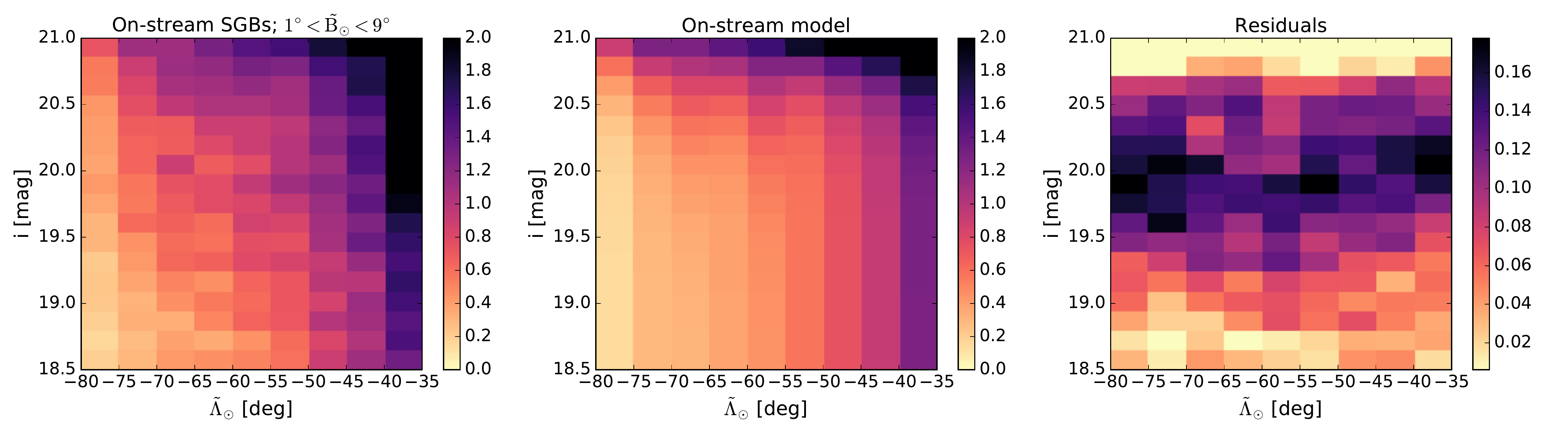}
  \caption{Same as Figure~\ref{trace_sgb}, but now for the SGB stars in the 
region between $1^{\circ} < \tilde{\rm B}_{\odot} < 9^{\circ}$, i.e. the 
so-called faint branch. At each $\tilde{\Lambda}_{\odot}$ bin, the residuals are 
vertically normalized. The diagrams are 9 $\times$ 17 pixels, and the residuals 
are smoothed with a Gaussian kernel with a FWHM of 0.7 pixels In this case, the 
signal of SGB stars is most prominent at $i \sim$ 20 mag, with a secondary peak 
at $i \sim$ 20.5 mag. In some $\tilde{\Lambda}_{\odot}$ bins, a small increase 
in the number of SGB stars is found at $i \sim$ 19 mag.}
    \label{trace_faint}
\end{figure*}
\begin{figure*}
  \centering
  \includegraphics[width=\textwidth]{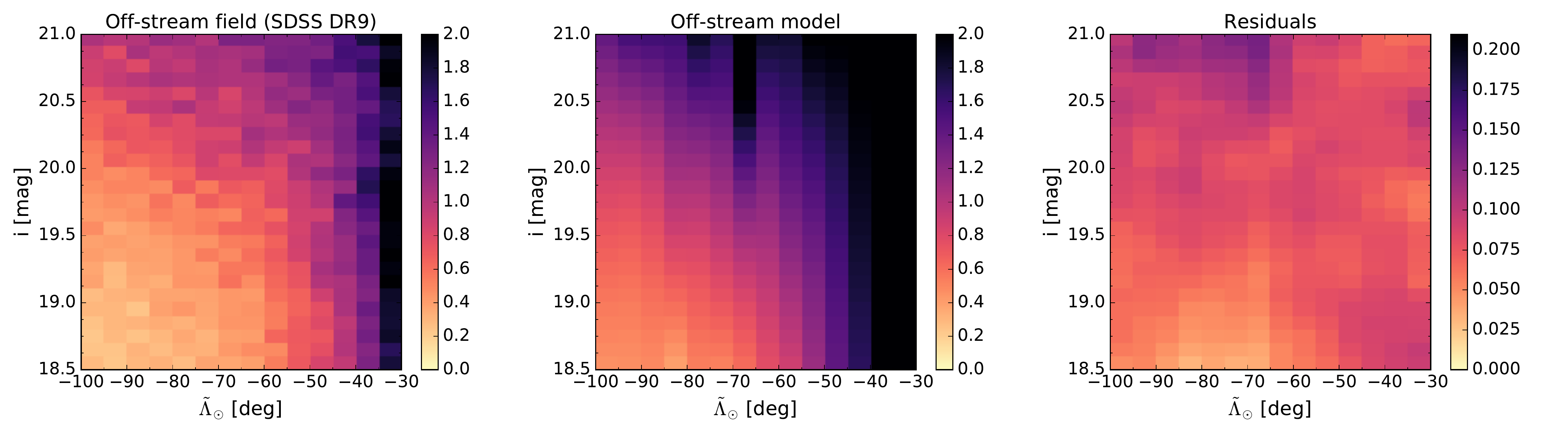}
  \caption{Same as Figure~\ref{trace_sgb}, but now for the SGB stars in the 
off-stream test field (covered by the SDSS DR9). The off-stream field (left), 
the model (center) and the residuals (right) are shown. At each 
$\tilde{\Lambda}_{\odot}$ bin, the residuals are vertically normalized and the 
residuals were smoothed using a Gaussian kernel with a FWHM of 0.7 pixels. No 
positive residuals can be traced in the magnitude range of interest, as expected 
for an area without any known halo sub-structure (with distances similar to 
those of the Sgr stream).}
  \label{background}
\end{figure*}

As a first diagnostic of the stellar populations in the Sgr stream as
viewed by the VST ATLAS, let us inspect the Hess diagram of the
portion of the trailing tail closest to the progenitor. More
precisely, the CMD density distribution is constructed for stars in a
region covering $-7^{\circ} < \tilde{\rm B}_{\odot} < 0^{\circ}$ and
$-50^{\circ} < \tilde{\Lambda}_{\odot} < -32^{\circ}$ (lower red box
in Figure~\ref{fig:stitch}). Assuming that the Milky Way is roughly
symmetric about the Galactic plane, we select an equivalent region
(i.e. the same range of Galactic longitude, but positive Galactic
latitude) in the North to be used as a representative of the
background Milky Way population. Given that ATLAS does not observe the
Northern sky, we take the photometry available from the SDSS DR9. Note
that the ATLAS and the SDSS photometric systems do not match exactly,
and therefore there may be magnitude offsets between the Hess diagrams,
of the order of 0.03 mag.

Figure~\ref{fig:hess}, from left to right, shows the Hess diagram for
the on-stream field, that for the mirror patch of the sky and the
difference of the two. The left and middle Hess diagrams were divided
by the area subtended on the sky by each region, and therefore each
bin represents the number of stars per magnitude per color, and per
square degree. The Sgr population is evident already in the on-stream
Hess diagram, even before the subtraction of the background. In the
background-subtracted diagram, the MS, the sub-giant and the red giant
branches can now be clearly seen. When the left panel of
Figure~\ref{fig:hess} is compared to the third panel of Figure 4 of
\citet{B06}, it is immediately apparent that the SGB portion of the
Hess diagram of the trailing tail in the ATLAS data is much broader than the
SGB of the leading tail in the SDSS. In fact, there is a striking
resemblance of the behavior of the SGB considered here and the split
SGB shown in the left panel of Figure 4 of \citet{B06}. In the North,
the double SGB arises due to the projection along the line of sight of
the two independent Sgr streams, the so-called branches A and C. We
conjecture that a similar situation is observed in this portion of the
trailing tail in the South: the possible existence of at least two,
vertically offset populations is betrayed by the fatness of the MSTO
and the thickness of the sub-giant and the red giant branches.

We have also examined the Hess diagram of the faint trailing tail, i.e. the 
portion of the sky with 1$^{\circ} < \tilde{\rm B}_{\odot} <$9$^{\circ}$ and 
$-50^{\circ} < \tilde{\Lambda}_{\odot} < -34^{\circ}$ (upper red box in 
Figure~\ref{fig:stitch}). Figure~\ref{fig:hess_faint} shows the on-stream Hess 
diagram, the off-stream Hess diagram, and the difference between the two. As in 
the previous case, the Hess diagrams are divided by the area covered by each 
region. The Sgr population is recovered as well, but it is much less prominent 
compared to the bright stream. Intriguingly, this branch of the stream ~-- 
studied earlier by \citet{K12} and \citet{S13} ~-- also does not possess a 
particularly tight SGB, thus hinting at multiple debris along the line of sight 
in the direction of the faint trailing Sgr branch.

To confirm the existence of the two populations offset in magnitude, the 
luminosity functions (LF), i.e. the one-dimensional slices for the on- and off-stream 
fields can be directly compared in Figure~\ref{fig:lf}. Based on the Hess 
diagram, we select SGB stars in the color range 0.52 $< (g-i) <$ 0.6 mag (dashed 
vertical lines in Figure~\ref{fig:hess} and \ref{fig:hess_faint}). According to 
Figure~\ref{fig:lf}, where the leftmost panel shows the on-stream LF (solid 
line) and the off-stream LF (dashed line, scaled to the ratio of areas between 
the two fields), the typically rather narrow SGB region is much broader than 
expected, spanning as much as 1 mag. Note that the steep rise of the luminosity 
function in the on-stream data from $i = 20.5$ mag onwards is mostly due to the 
stream MS stars. The difference between the luminosity functions (left middle 
panel) shows two distinct peaks at $i =$ 19.6 and 20.0 mag, with the faintest 
one being the most prominent of the two. Given that the SGB peaking at the 
fainter magnitude contains more stars, we choose to designate it SGB 1, and its 
counterpart, peaking $\sim0.6$ mag brighter, SGB 2. Using the off-stream LF 
as a model of the background, the significance of SGB 1 and SGB 2 is 6.1 and 
5.3$\sigma$, respectively.

In the case of the faint Sgr stream, the on-stream LF (right middle
panel) also shows a broad SGB peak around $i\sim 19.8$ mag. The
subtraction of the off-stream field (from SDSS data) shows the
presence of three peaks at $i \sim$ 19.3, 19.9 and 20.4 mag. However,
in this case, the number counts are much less than for the Sgr bright
stream. In fact, the significance of these peaks is much lower:
  SGB 1a has 3.0$\sigma$ of significance, while the secondary peak
  (dubbed SGB 1b) has only 1.5$\sigma$. SGB 2 also has a low
  significance, namely $\sim 2\sigma$. Despite being hardly
  statistically significant, the detections are also recovered with
  other tracers (see Section~\ref{sec:bhbs}), which suggests they
  might indeed be real substructures, but less prominent than the two
  streams found in the bright trailing arm (above the Sgr's plane).

The presence of two peaks in the SGB luminosity function of the bright Sgr 
stream (i.e., $\tilde{\rm B}_{\odot} < 0^{\circ}$), separated by about 0.6 mag, 
could be either due to two distinct Sgr tidal debris overdensities at two 
different distances along the line of sight, or, alternatively, due to a complex 
stellar population in the Sgr stream, albeit all at the same distance. Another 
possibility that cannot be ruled out is that one of the two detected populations 
corresponds to an entirely new stream, not connected to the Sgr dwarf in any 
way. In what follows, we shall trace the position of the stars in the two SGB 
peaks as a function of the longitude along the Sgr coordinate system, in an 
attempt to elucidate the genesis of the split SGB. The same analysis is 
performed for the faint Sgr tail. However, as mentioned before, the core of this 
work is the analysis of the SGB detections on the bright Sgr tail; while the 
analysis of the faint one should be considered as a first attempt to trace the 
density evolution of this less populated signature.

\section{Stream tomography}
\label{sec:tomography}

\begin{figure*}
  \centering
  \includegraphics[width=0.45\textwidth]{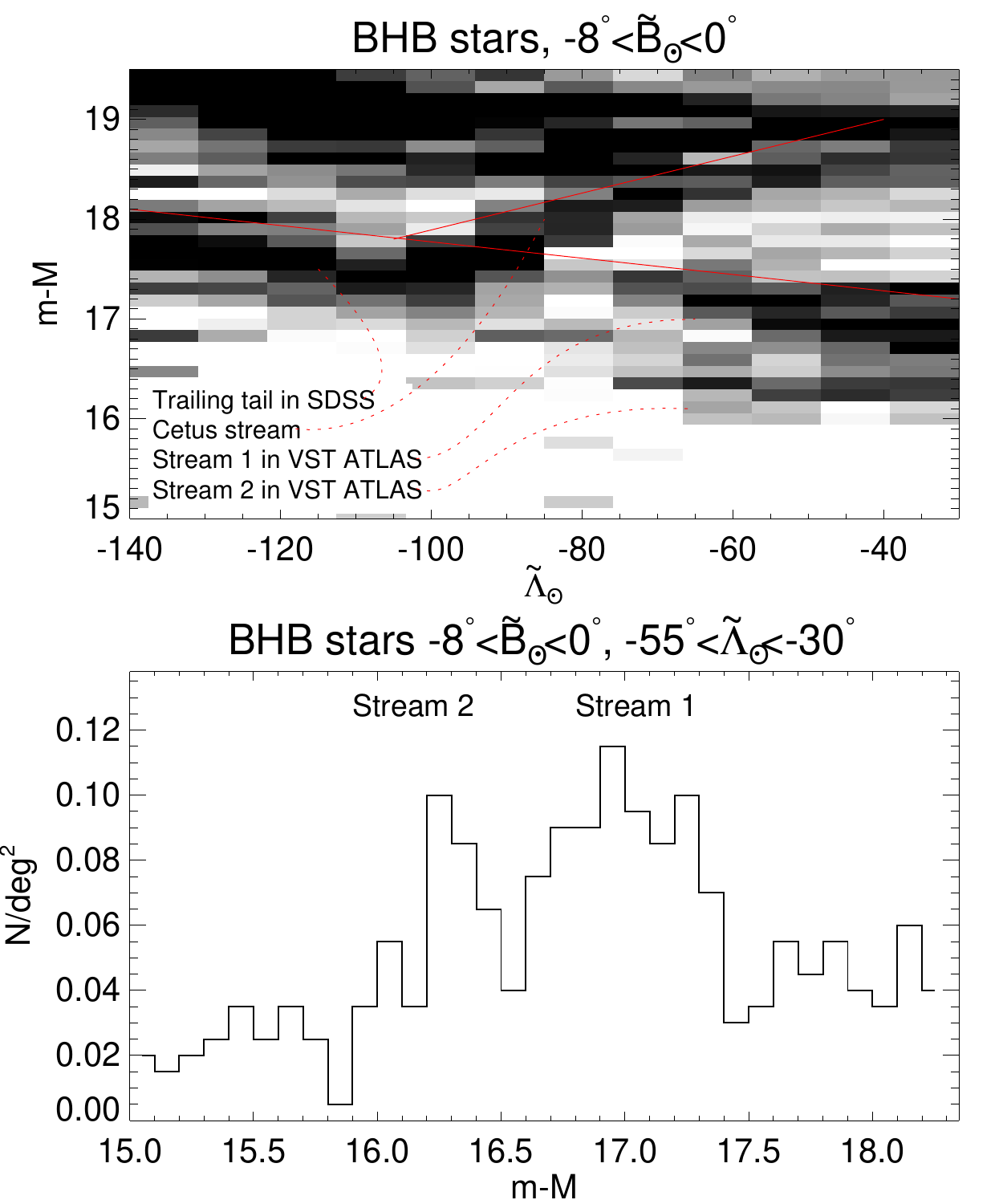}
  \includegraphics[width=0.45\textwidth]{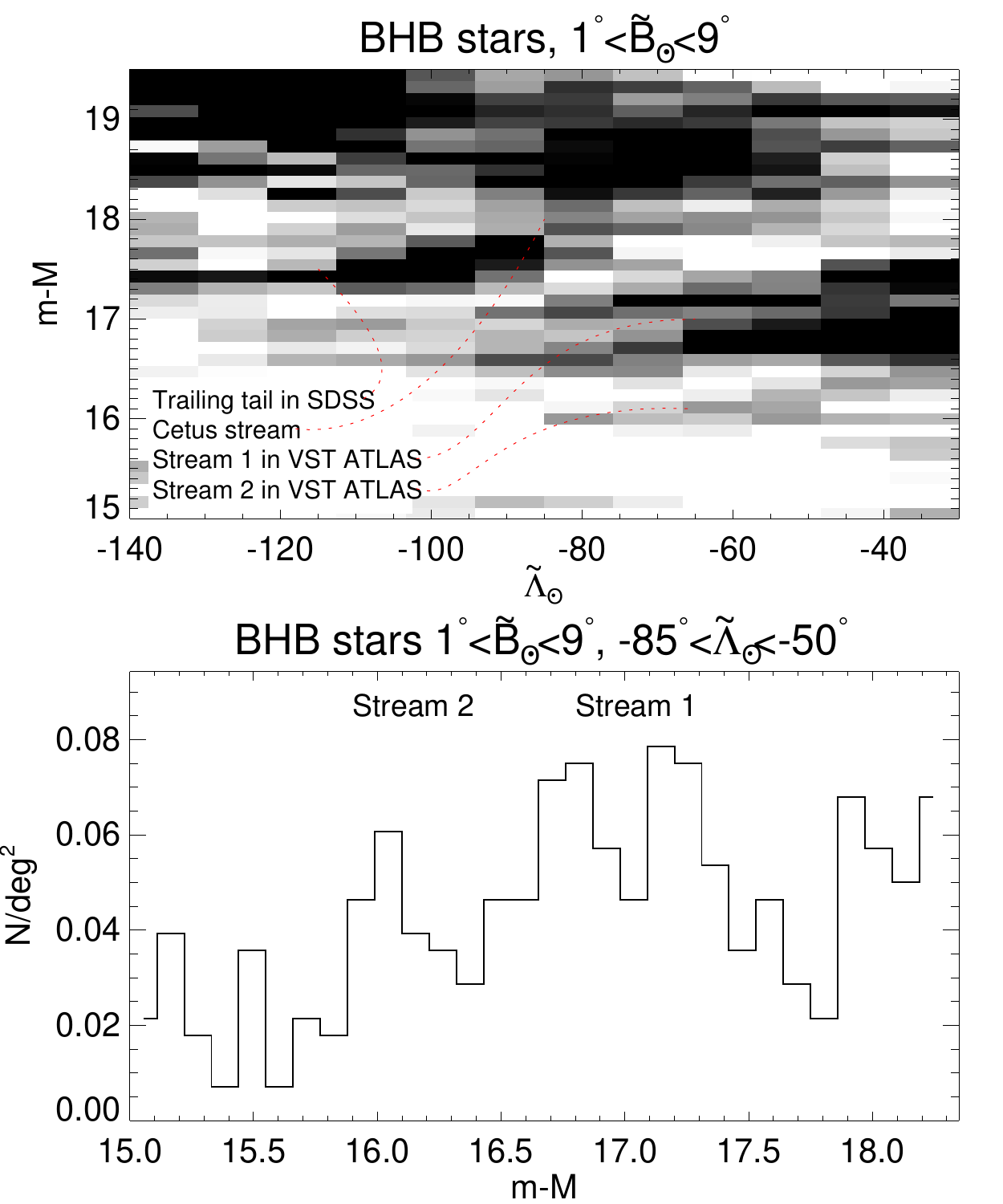}
  \caption{Line-of-sight distribution of BHB candidate stars. {\it
      Top:} Greyscale density of the BHB candidate stars in the plane
    of distance modulus vs. Sgr stream longitude
    $\tilde{\Lambda}_{\odot}$ for the bight (left) and the faint
    (right) trailing arm branches. The darker regions correspond to an
    enhanced stellar density. Note that a combination of BHB candidate
    stars from both SDSS and VST ATLAS datasets is shown here. The
    main structures, including the Sgr and the Cetus Polar Stream
    debris, are marked. The two solid lines show the distance
      gradient of the Sgr and the Cetus streams. {\it Bottom:} Slices
    through the 2D BHB density plots shown in the Top panel for a
    small range of $\tilde{\Lambda}_{\odot}$ (as indicated in the
    title of each panel). Peaks corresponding to different
    line-of-sight components of the trailing arm are clearly visible
    (see main text for a detailed discussion).}
    \label{bhbs}
\end{figure*}

\subsection{Exposing the Sagittarius SGB}

Figure~\ref{trace_sgb} shows the density of the candidate SGB stars
with $-7^{\circ} < \tilde{\rm B}_{\odot} < 0^{\circ}$ in the space of
$i$ magnitude and $\tilde{\Lambda}_{\odot}$. Here, a much larger
extent of the Sgr longitude is studied, i.e. $-100^{\circ} <
\tilde{\Lambda}_{\odot} < -30^{\circ}$, corresponding approximately to
the entire additional stream coverage supplied by the ATLAS survey. In
order to trace the variation (if any) in the magnitude of the two SGB
peaks along the Sgr longitude, the stream region is divided into 14
segments, each 5 degrees long, while keeping the vertical bins at 0.1
mag. In the left panel of the Figure, two tendencies are immediately
apparent: the steep ``vertical'' rise of the star counts, i.e. larger
densities at fainter magnitudes due to i) the increased halo
contribution with growing volume probed and ii) increased contribution
from the MS stars in the Sgr stream; as well as the ``horizontal''
density increase to lower $|\tilde{\Lambda}_{\odot}|$ due to the drop
in the Galactic latitude. Note that the abrupt decrease in the density
at $\tilde{\Lambda}_{\odot} > -40^{\circ}$ is due to the
incompleteness of the ATLAS footprint. It is evident that the
stream SGB signal is buried under the strong density gradients in both
the magnitude and longitude dimensions.

To reveal the Sgr stream SGB, rather than trying to identify an
appropriate patch of the sky for comparison, we put forward the
following simple model of the background star counts. In each bin of
$\tilde{\Lambda}_{\odot}$, the luminosity function is approximated
with an exponential profile $N(mag) = A\exp(B\times mag)+C$. We allow
the model parameters $A, B$ and $C$ to vary from bin to bin along
$\tilde{\Lambda}_{\odot}$, and mask out the magnitude range of
interest, i.e. pixels with $18.8 < i < 20.2$ (see
Figure~\ref{fig:lf}). The middle panel of Figure~\ref{trace_sgb} shows
the variation of the model luminosity function obtained.

The right panel of Figure~\ref{trace_sgb} shows the distribution of
the residuals after subtraction of the model described above from the
data. In order to improve the visibility of the overdensities along
each line of sight, this 2D histogram is vertically normalized,
i.e. the counts in each $\tilde{\Lambda}_{\odot}$ bin are divided by
the total in each column and the binsize. The SGB signal is now
clearly discernible between $i=19$ and $i=20.5$.  From right to left,
i.e. from lower $|\tilde{\Lambda}_{\odot}|$ values to higher, a
shallow apparent magnitude gradient is visible, with the stream
detectable at slightly fainter magnitudes around
$\tilde{\Lambda}_{\odot}=-100^{\circ}$ compared to
$\tilde{\Lambda}_{\odot}=-30^{\circ}$. The SGB magnitude gradient is
due to the slow evolution of the heliocentric distance to the Sgr
stream with $\tilde{\Lambda}_{\odot}$ as measured previously
\citep[see e.g.][]{K12,S13} and in agreement with the model of the
dwarf disruption (see Section~\ref{sec:dandc}). Additionally, the
width of the SGB band appears to evolve with the Sgr
longitude. Finally, in agreement with the analysis reported in the
previous Section, in the range
$-60^{\circ}<\tilde{\Lambda}_{\odot}<-30^{\circ}$ two individual SGB
components are visible. These are separated by $\sim0.6$ mag at
$\tilde{\Lambda}_{\odot} \sim -30^{\circ}$ and (nearly) merge together
at $\tilde{\Lambda}_{\odot}\sim -70^{\circ}$.

Analogously, Figure~\ref{trace_faint} shows the behavior of the
candidate SGB stars with $1^{\circ} < \tilde{\rm B}_{\odot} <
9^{\circ}$, i.e. in the area corresponding to the faint Sgr branch. As
in the previous Figure, the panels show the on-stream field (left),
the on-stream model (middle) and the vertically-normalized residuals
(right). The stream region was divided into 5 degrees long bins while
the vertical bins were increased up to 0.15 mag, but the
  vertically-normalized residuals are also divided by the binsize, in
  order to reach the same signal level as in
  Figure~\ref{trace_sgb}. In this case, to construct the model the
bins in the range 19.0 $< i <$ 20.5 mag were excluded.  The residuals
show a strong detection of an overdensity at similar apparent
magnitude as compared to that in the bright branch. However, here the
SGB signal displays a slightly different behavior with
$\tilde{\Lambda}_{\odot}$, and, overall, the line-of-sight
distribution of the SGB stars is more complicated, varying from narrow
to broad, and in some directions showing what can be interpreted as
two or even three distinct peaks.

In the following sub-section we measure the positions of the SGB peaks along the 
line of sight and convert apparent magnitudes into heliocentric distances. 
Before that, however, we perform a simple test to ensure the robustness of 
the analysis. As a sanity check, we apply the luminosity function modeling 
analysis described above to an SDSS field with equatorial coordinates 
(190$^{\circ}$, 47$^{\circ}$), i.e. a location in the Galaxy where no obvious 
stellar halo overdensities have been detected to
date in the distance range of interest. Figure~\ref{background} shows the 
density of the candidate SGB stars in the SDSS selected field with the same 
arrangement of panels as in the previous two figures. Interestingly, the density 
variations in the first two panels (data and model) appear to resemble those in 
the bright and faint streams shown in Figures~\ref{trace_sgb} and 
~\ref{trace_faint}. However, reassuringly, in the third panel no strong positive 
residuals are visible, thus giving us confidence that the signal revealed in the 
previous two figures is genuine.

\begin{table}
 \centering
 \caption{Distance modulus to the trailing SGB 1 and 2 peaks along the 
bright branch of the Sgr stream.}
 \label{tab:table1}
 \begin{tabular}{lcccc}
  \hline
  \hline  
  $\tilde{\Lambda}_{\odot}$      & m-M (SGB 1)    & $\sigma$(m-M) & m-M (SGB 2) & 
$\sigma$(m-M) \\
  ($^{\circ}$)    & (mag)        & (mag)         & (mag)     & (mag)         \\
  \hline
        --97.5    & 17.20        & 0.10         & 16.74     & 0.15           \\
        --92.5    & 17.10        & 0.07         & 16.71     & 0.15           \\
        --87.5    & 17.09        & 0.06         & 16.52     & 0.16           \\	
        --82.5    & 17.01        & 0.10         & 16.85     & 0.07           \\	
        --77.5    & 17.05        & 0.04         &           &                \\	
        --72.5    & 16.87        & 0.11         & 16.71     & 0.07           \\	
        --67.5    & 16.92        & 0.06         & 16.48     & 0.12           \\	
        --62.5    & 16.78        & 0.15         & 16.63     & 0.12           \\	
        --57.5    & 16.80        & 0.05         & 16.31     & 0.06           \\	
        --52.5    & 16.84        & 0.05         & 16.44     & 0.10           \\	
        --47.5    & 16.82        & 0.16         & 16.52     & 0.12           \\	
        --42.5    & 16.87        & 0.03         & 16.28     & 0.03           \\	
        --37.5    & 16.96        & 0.13         & 16.37     & 0.17           \\	
        --32.5    & 16.80        & 0.07         & 16.29     & 0.11           \\	
  \hline
 \end{tabular}
\end{table}

\begin{table*}
 \centering
 \caption{Distance modulus to the trailing SGB 1 and SGB 2 peaks in the SGB 
luminosity function along the Sgr faint stream}
 \label{tab:faint_dist}
 \begin{tabular}{lcccccc}
  \hline
  \hline  
  $\tilde{\Lambda}_{\odot}$  & m-M (SGB 1a)    & $\sigma$(m-M) & m-M (SGB 1b) & 
$\sigma$(m-M) & m-M (SGB 2) & $\sigma$(m-M) \\
  ($^{\circ}$)                    & (mag)        & (mag)         & (mag)     & 
(mag)  & (mag) & (mag)       \\
  \hline
       --77.5                     & 16.84 & 0.06 & 17.48 & 0.13 & 15.97 & 0.09 
\\
       --75.5                     & 17.00 & 0.14 & 17.37 & 0.11 &       &  \\
       --67.5                     & 16.88 & 0.12 & 17.48 & 0.06 &       &  \\
       --62.5                     & 16.69 & 0.10 & 17.34 & 0.09 & 15.99 & 0.16 
\\
       --57.5                     & 16.56 & 0.08 & 17.51 & 0.25 & 15.79 & 0.08 
\\
       --52.5                     & 16.83 & 0.13 & 17.45 & 0.11 & 15.84 & 0.19 
\\
       --47.5                     & 16.71 & 0.12 & 17.39 & 0.06 & 15.83 & 0.09 
\\
       --42.5                     & 16.89 & 0.07 & 17.47 & 0.03 &       &  \\
       --37.5                     & 16.94 & 0.05 & 17.52 & 0.08 & 15.97 & 0.08 
\\
  \hline
 \end{tabular}
\end{table*}

\subsection{Distances to the Sgr trailing tail with SGB stars}
\label{sec:dist}

With the Sgr SGB signal now cleansed of the contamination, we fit a
Gaussian mixture model to the residual density distributions shown in
the right panels of Figures ~\ref{trace_sgb} and ~\ref{trace_faint} to
obtain the centroids of the SGB peaks. To get an idea of the
systematic uncertainty of our peak centroiding procedure, we change
the magnitude bins by $\pm 50$\% and re-measure the peak
positions. The difference in the positions of each peak obtained with
three different magnitude bin sizes is taken as representative of
the systematic error in the method. At each Sgr longitude, the error
bars for each peak are the sum (in quadrature) of the random errors on
the peak determination from the Gaussian fit and the systematic error
associated to the $i$ magnitude bin size. The same procedure is
adopted to recover the centroids of the peaks in the Sgr faint stream.

 The errors in distance modulus listed do not represent the
  possible dispersion in distance modulus along the line of sight. For
  each of the SGB detections, the dispersion of the distribution could
  be gleaned from the $\sigma$ of the Gaussian fit. For the faint arm
  detections, the three peaks have similar dispersions, generally
  lower than 0.2 mag. In the case of the detection in the bright arm,
  the dispersions are of the order of 0.3 mag. There is no evident
  trend of the peak as a function of $\tilde{\Lambda_{\odot}}$,
  neither the heliocentric distance with the associated dispersion
  along the line of sight. The same is not true for the RR Lyrae stars
  (see Section~\ref{sec:rrl}), which reveal a marked difference in
  heliocentric distance dispersion at different
  $\tilde{\Lambda_{\odot}}$.

The absolute distance calibration for the SGB peaks thus measured can be 
obtained using previous detections of the stream at the same 
$\tilde{\Lambda}_{\odot}$, based on tracers with reliable distances. The 
distances to the Sgr debris at $\tilde{\Lambda}_{\odot} = -97.5^{\circ}$ and $-92.5^{\circ}$ 
were previously measured by \cite{K12}, using the RC stars\footnote{Note that we 
offset the distances by a small amount to improve agreement with the measurement 
based on the RR Lyrae stars as described in Section~\ref{sec:rrl}.}. Given that 
the SGB 1 is more luminous than the SGB 2, we take it as the main branch of the 
Sgr trailing tail and chose to match its distance to that previously detected 
using the RC. Tables~\ref{tab:table1} and ~\ref{tab:faint_dist} present the 
distance modulus for the SGB detections in both Sgr trailing branches, for each 
$\Lambda$ bin, and the total errors calculated as described above.

\subsection{Sgr trailing debris with BHB stars}\label{sec:bhbs}

The Sgr trailing debris sub-structure discovered above ought to be
visible with tracers other than SGB stars. Consequently, we slice
through the stream using BHB stars. The latter have long been the
working horse of Galactic stellar halo studies thanks to their
unique properties: low levels of contamination, well-calibrated
absolute magnitudes and robust identification by means of multi-band
photometry only \citep[see
  e.g.][]{Yanny2000,Clewley2002,Newberg2003,Sirko2004,Xue2008,Bell2010,Deason2011,
  Ruhland2011,Deason2012,Deason2014,B14,Be2016}. As mentioned earlier,
VST ATLAS provides $ugriz$ measurements, therefore BHB stars can be
selected with the $u-g$, $g-r$ cuts similar to those normally used for
the SDSS photometry, as explained in the literature referenced
above. In practice, given that the VST ATLAS $u$ band does not match
the SDSS $u$ exactly, the BHB selection boxes differ slightly between
the two surveys. More precisely, we use the simplified version of the
BHB color cuts, i.e. for the SDSS: $ 1 < u-g < 1.5,~-0.35 < g-r <
-0.05$, and $0.9 < u-g < 1.3, ~-0.35 < g-r <-0.05$ for the VST
ATLAS. We assign distances to the candidate BHB stars using the
absolute magnitude calibration of \citet{Be2016} which is almost
identical (in terms of the absolute magnitude behavior as a function
of the $g-r$ color) to that of \citet{Deason2011}.

The top row of Figure~\ref{bhbs} gives the density of the candidate
BHB stars along the line of sight as a function of
$\tilde{\Lambda}_{\odot}$. Here we extend the range of Sgr longitude
further away from the progenitor to be able to reproduce the previous
Sgr debris detections from e.g. \citet{K12}. Bear in mind that the BHB
selection employed aims to maximize completeness of BHB stars, at the
cost of an higher (compared to more sophisticated color-color
selection boxes) contamination. Typical contaminants for us to
consider are the Blue Stragglers (BS) and MSTO stars. These, however,
are fainter by $\sim 1.7$ mag and can be easily identified as a broad
``shadow'' to the narrow BHB sequence.  The top left panel of the
Figure explores the range of $-8^{\circ} < \tilde{\rm B}_{\odot} <
0^{\circ}$, appropriate for the bright branch of the Sgr trailing
tail. The counts in each column have been normalized by the column's
total to emphasize the strongest overdensities at each
$\tilde{\Lambda}_{\odot}$. The two solid lines show the distance
  gradient of the Sgr and Cetus stream. From $\tilde{\Lambda}_{\odot}
  = -140^{\circ}$ to $-30^{\circ}$ the BHBs associated with the Sgr
  stream proceed from m-M$\sim$17.6 mag to $\sim$17.0 mag. On the
  contrary, the heliocentric distance of the BHBs in the Cetus stream
  increases with $\tilde{\Lambda}_{\odot}$. The streams reach the same
  distance modulus at $\tilde{\Lambda}_{\odot} = -100^{\circ}$. In the
  region studied here, and in particular near the Sgr dwarf core, the
  Cetus stream is detected more than 1 mag further away than the
  Sgr. The strongly different distant gradients discard the
  possibility that some of our detections could belong to the CPS.

Starting at large distances from the Sgr dwarf, at
$-140^{\circ}<\tilde{\Lambda}_{\odot} < -100^{\circ}$, the narrow dark
stripe corresponding to the overdensity of BHB stars can be seen at $17 <
m-M < 18$. It is followed by a broader BS ``shadow'' at $m-M>19$. Both
BHB and BS sequences show a shallow distance gradient with
$\tilde{\Lambda}_{\odot}$. Around
$-100^{\circ}<\tilde{\Lambda}_{\odot} < -70^{\circ}$, the Sgr stream
overlaps with another stellar halo sub-structure, the Cetus Polar
Stream \citep[CPS;][]{Newberg2009}. As demonstrated in \citet{K12}, on
the sky, the Cetus stream runs at an angle with respect to the track
of the Sgr trailing tail. Along the line of sight, the Cetus stream
follows a distance gradient opposite to that of the Sgr trailing
debris. At $-100^{\circ}<\tilde{\Lambda}_{\odot} < -30^{\circ}$, the
Sgr trailing tail continues along the distance gradient seen at 
higher $|\tilde{\Lambda}_{\odot}|$, but is now clearly much broader
than its continuation further down the stream or compared to Cetus. In
fact, hints of two distinct sequences can be discerned.

To scrutinize the line-of-sight sub-structure of the Sgr trailing
tail, the lower row of Figure~\ref{bhbs} shows 1D slices through the
2D density distributions given in the top row. In particular, the
bottom left panel displays the histogram of stars in the bright
branch, i.e. at negative $\tilde{\rm B}_{\odot}$ in the restricted
range of $-55^{\circ}<\tilde{\Lambda}_{\odot}<-30^{\circ}$. Two
distinct bumps are noticeable: a narrow one with a peak at
$m-M\sim16.3$ and a broad one with the maximum at $m-M\sim
16.9$. These appear to be in perfect agreement with the distance
measurements based on the SGB analysis reported in
Table~\ref{tab:table1} for the same range of
$\tilde{\Lambda}_{\odot}$. Stream 1 contains $\sim$150 stars,
  while the ``background'' of the distribution has $\sim$60 stars,
  consistent with a peak significance $>$10$\sigma$. For Stream 2 the
  significance is $\sim$5$\sigma$, with an estimated background of 24
  stars and 26 BHB stars in excess.

Similarly, the right column of Figure~\ref{bhbs} presents the behavior
of the density of the candidate BHB stars in the faint branch of the
trailing tail.  Here, features similar to those in the left column are
observable, albeit with some curious differences. For example, only a
small portion of the Cetus stream can be seen at
$-100^{\circ}<\tilde{\Lambda}_{\odot}<-80^{\circ}$. This is due to the
misalignment between CPS and Sgr, with the former moving down to lower
declinations. Note that the counterparts of both SGB 1 (more luminous
component, at $m-M\sim 17$) and SGB 2 (less prominent overdensity, at
$m-M\sim16$) can be discerned. Interestingly, Stream 1 appears to
break up into further two sub-structures, again in agreement with the
complex signal in the SGB residuals map presented above. The
  significance of the two sub-structures in Stream 1 is
  $\sim$3$\sigma$ each, i.e. twice as significant as their SGB
  counterparts (see Section~\ref{cmd}). In particular, the peak at
  $m-M\sim17.2$ mag has 43 stars while there are 26 in the background.
  The nearest component, at $m-M\sim16.8$ mag, has 57 BHBs (with
  $\sim$38 stars in the estimated background). Considering the
  contribution of both peaks, as one single detection, the number of
  BHBs is 150, with a background level of $\sim$75 stars, giving an
  overall significance greater than 10$\sigma$. The stream 2 has a
  4$\sigma$ significance (20 stars in excess, over an estimated
  background of 20 BHBs).
  
Overall, the BHB stars fully corroborate the picture established
earlier with the SGB tracers: the trailing debris appears to be
bifurcated into two components along the line of sight. Although,
  he bright Sgr trailing arm detections appear to be much more
  significant compared to the ones found at the faint arm. Note,
however, that without kinematic information it is impossible to
establish with 100\% certainty that the brighter BHBs have indeed
originated from the Sgr dwarf. Perhaps, these could belong to a
different halo sub-structure, unrelated to Sagittarius. To clarify
further the nature of the BHB stars with lower $m-M$, their position
on the sky is studied in Section~\ref{sec:3rd}.

\subsection{Sgr trailing tail with CRTS RR Lyrae}\label{sec:rrl}

RR Lyrae (RRL) are old (at least 10 Gyr old) and metal-poor pulsating
stars. RRL are excellent distance indicators and indispensable
tracers of sub-structure in the halo \citep[see e.g.,][and references
  therein]{C09, C15}. However, they are much less numerous compared
to other stellar tracers, such as MSTO, SGB and often even BHB stars, and
generally require variability surveys to be identified. RRL data from
previous wide-field and dedicated surveys have been used to trace the
Sgr stream. These include the SDSS Stripe 82 \citep{I00, W09}, 
QUEST \citep{V05}, and the SEKBO survey \citep{P09}. The datasets
above have mostly focused on the Northern sky, at declinations
generally higher than $-10^{\circ}$, thus missing the portion of the
Sgr stream closest to the progenitor.

\begin{table}
 \centering
 \caption{Distance modulus peaks of RRL stars along Sgr trailing tail}
 \label{tab:rrl_distance}
 \begin{tabular}{lcccc}
  \hline
  \hline  
  $\tilde{\Lambda}_{\odot}$ Sgr   & m-M (RRL 1)    & $\sigma$(m-M) & m-M (RRL 2) & 
$\sigma$(m-M)  \\
  ($^{\circ}$)                    & (mag)        & (mag)         & (mag)     & 
(mag)         \\
  \hline
       --44.38                    & 16.87 & 0.08 & 16.40 & 0.19 \\
       --33.13                    & 16.82 & 0.14 & 16.28 & 0.27 \\
       --21.88                    & 16.90 & 0.04 &       &      \\
       --10.63                    & 17.11 & 0.02 &       &      \\
  \hline
 \end{tabular}
\end{table}

\begin{figure}
\includegraphics[width=0.45\textwidth]{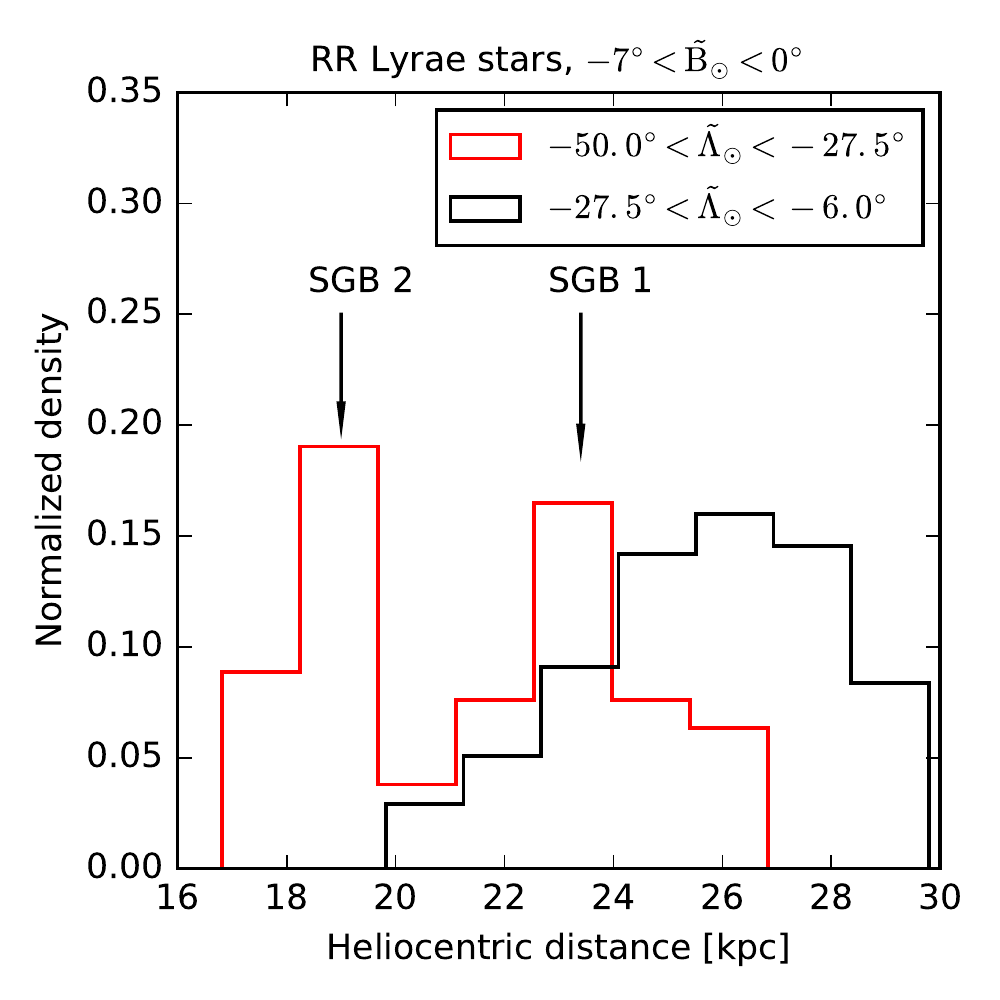}
\caption{Heliocentric distances for RRL stars in Sgr\,1, restricted to the Sgr 
bright arm region (i.e., $-7^{\circ} < \tilde{\rm B}_{\odot} < 0^{\circ}$). The 
red  histogram corresponds to RRL stars located at $-50^{\circ} < 
\tilde{\Lambda}_{\odot} < -27.5^{\circ}$ and the black histogram to those 
located at $-27.5^{\circ} < \tilde{\Lambda}_{\odot} < -6.0^{\circ}$. The 
distribution of RRL stars located near the edge of VST detections, further away 
from Sgr dwarf, is bimodal, with two peaks consistent with the detections of SGB 
and BHB stars from VST ATLAS survey. For the innermost region, where VST does 
not observe, the RRL stars have a distribution consistent with only one main 
peak, but wider than the previous detections.}
     \label{fig:rrl_peaks}
 \end{figure}

\begin{figure*}
\includegraphics[width=0.48\textwidth]{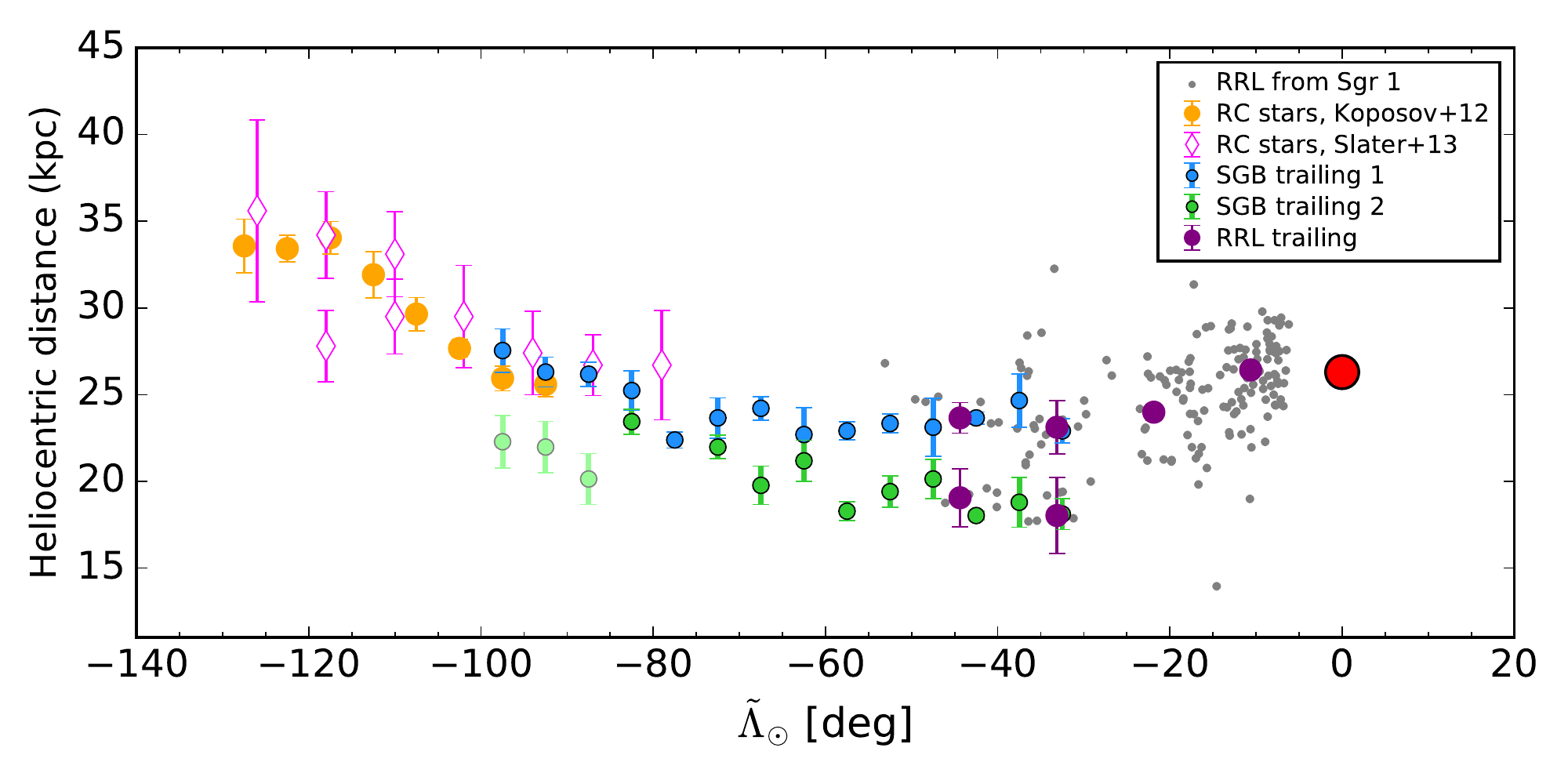}
\includegraphics[width=0.48\textwidth]{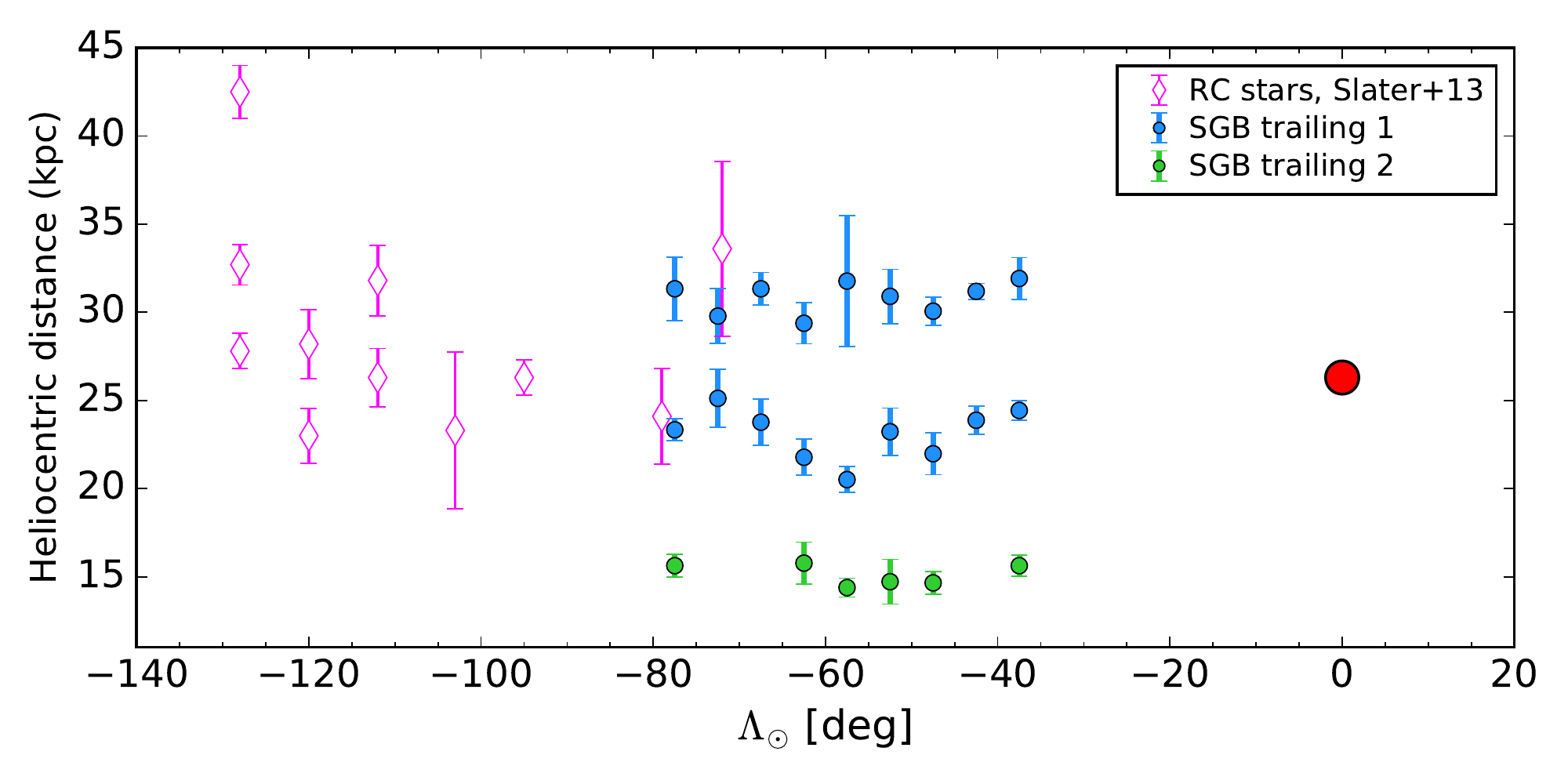}
\caption{{\it Left:} Heliocentric distances as a function of the longitude along the stream. 
Yellow circles correspond to the RC measurements by \protect\cite{K12}, while 
open pink rhombs are the RC detections of \protect\cite{S13} using the 
Pan-STARRS data. The SGB data-points obtained in this work are marked with blue 
(trailing 1) and green (trailing 2) circles. Peaks in the RRL distribution 
\citep[using the Sgr~1 sample of][]{T15} are marked 
with purple circles, while the individual RRL stars from Sgr~1 are marked 
with grey points. The position of the Sgr dwarf, at $\tilde{\Lambda}_{\odot}$ = 
0$^{\circ}$, and heliocentric distance of 26.3 kpc, is marked with a filled red 
circle. {\it Right:} Heliocentric distances as a function of 
$\tilde{\Lambda}_{\odot}$ for the faint branch of the trailing arm (located at 
$\tilde{\rm B}_{\odot} > 0^{\circ}$). Open diamonds correspond to the RC 
detections from \protect\cite{S13}. Measurements based on the SGB stars are 
marked with blue (trailing 1) and green (trailing 2) circles. The main stream, 
trailing 1, splits into two parallel sequences, separated by $\sim$5 kpc in 
heliocentric distance.}
     \label{dist}
 \end{figure*}

The Catalina Real-Time Transient Survey \citep[CRTS;][]{L03, D09} has
filled the gap in the RRL map of the Milky Way, detecting variable stars
across the Southern hemisphere, reaching as low as $-75^{\circ}$ in
declination. Recently, \cite{T15} identified several promising
overdensities in the Southern sky using RRL stars from CRTS. The most
significant of these overdensity candidates is dubbed Sgr~1, with a
$>15\sigma$ significance. Based on its 3D position, Sgr~1 is very
likely a part of the trailing tail of the Sgr dwarf in the South. The
Sgr~1 sub-structure contains 327 RRL stars, assuming a model for the
expected number of RRL in the halo at that position on the sky
\citep[for details, see][]{T15}. Clearly, with a substantial tally of
327 RRL stars, Sgr 1 more than fills in the gap in the Sgr stream
coverage in the vicinity of the progenitor.

From the 327 RRL stars originally identified in the Sgr~1
overdensity, we selected only those confined to the plane of the
bright trailing arm of Sgr, i.e. between $-7^{\circ} < \tilde{\rm
  B}_{\odot} < 0^{\circ}$. This spatial selection reduces the the
sample size to 250 RRL, spanning the longitude of the Sgr stream from
$\tilde{\Lambda}_{\odot} = -50^{\circ}$ to $\tilde{\Lambda}_{\odot}
=-6^{\circ}$. Unfortunately, because of the spatial limits of the
survey, there are only 42 RRL located at positive $\tilde{\rm
  B}_{\odot}$, thus making it unfeasible to track down the peaks for the
faint Sgr branch. Therefore, the RRL distribution is only analyzed here for
the bright branch of the Sgr trailing arm.

To trace the distance gradient in RRL stars, the length of the Sgr~1 
overdensity was divided into 4 bins, each 11.25$^{\circ}$ wide, similar to the SGB 
analysis. The wider $\tilde{\Lambda}_{\odot}$ bins are chosen to ensure enough 
RRL stars in each bin. The RRL number density along the stream shows a strong 
gradient: indeed, the bins further away from the main body of Sgr are less 
populated, by a factor of two, compared to the two innermost bins (where there 
are more than 70 RRL stars per bin). For each $\tilde{\Lambda}_{\odot}$ bin, the 
heliocentric distance distribution was modeled with either one or two Gaussians. 
The heliocentric distances come directly from the catalogue published by 
\cite{T15}, and they are accurate to $\sim$10\% \citep[depending on the 
magnitude of the RRL, the distance uncertainty is between 7\% and 12\%, 
see][]{D13}. We detect two peaks in the heliocentric distance distribution of 
RRL in Sgr~1 in the first two bins, namely at $\tilde{\Lambda}_{\odot} \sim 
-40^{\circ}$ and $\tilde{\Lambda}_{\odot} \sim -33^{\circ}$. In the two bins 
closest to the progenitor, the distance distribution of RRL is consistent with 
one broad peak which was modeled with a single Gaussian. 
Figure~\ref{fig:rrl_peaks} shows the heliocentric distance distribution for RRL 
stars in Sgr~1 located at $-50.0^{\circ} < \tilde{\Lambda}_{\odot} < 
-27.5^{\circ}$ (red histogram) and at $-27.5^{\circ} < \tilde{\Lambda}_{\odot} 
< -6.0^{\circ}$ (dotted line histogram). RRL stars further away from the Sgr 
dwarf (i.e., at higher $| \tilde{\Lambda}_{\odot} |$) show a clear bimodality, 
with two peaks located at $\sim$19.3 and 23.4 kpc, nicely connecting the two 
peaks found using SGB and BHB stars. Closer to the dwarf, the distribution peaks 
at $\sim$26.0 kpc, with a wider range of distances. This suggests that most 
probably the two peaks forming the fork get merged as they are going towards the 
actual position of the Sgr dwarf nucleus. The gap between both distributions, at 
$\sim$20 kpc, is mostly due to selection effects as the sample of RRL stars from 
Sgr\,1 was selected based on the local overdensity of RRL stars with respect to 
the density of RRL stars in the halo \citep{T15}. In the figure, we adopted the 
same naming convention for the two peaks with most distant dubbed RRL 1 
while the nearest one is RRL 2. 

As the RRL stars are the most trustworthy distance indicators 
at our disposal, we can tighten up the SGB absolute distance scale. More 
precisely, using the RC distances from \citet{K12} as our reference point (see 
Section~\ref{sec:dist}), there appears to be a small offset of $\sim$0.3 kpc compared to the 
distances of RRL at $\tilde{\Lambda}_{\odot} = -45^{\circ}$. Therefore, we 
choose to match the SGB Sgr stream distances at $\tilde{\Lambda}_{\odot} 
=-44^{\circ}$ to those dictated by the RRL, i.e. 23.7 kpc. Indeed, an 
impressive match (as can be judged from Fig.~\ref{dist} below) between the two 
overdensities detected with SGB and RRL stars obtains as a result, when we 
offset the distances at $\tilde{\Lambda}_{\odot} = -97.5$ and $-92.5$ by 
0.4 and 2 kpc, respectively, as compared to the \citet{K12} scale. The distance 
modulus to each of the peaks detected as well as the errors (derived as in the 
case of the SGB stars) are reported in Table~\ref{tab:rrl_distance}.

   \section{Discussion and Conclusions}\label{sec:dandc}
  \subsection{Distances along the Sgr stream}
 Figure~\ref{dist} presents the heliocentric distance evolution along
the Sgr trailing arm as a function of $\tilde{\Lambda}_{\odot}$, as
measured with different stellar tracers. Included in the Figure are
the previous detections, such as the RC measurements of \cite{K12}
shown as filled orange circles, as well as the measurements based on the
SGB stars (green and blue circles) carried out here. The left panel of
the Figure shows the distance measurements for the bright component of
the trailing arm, i.e. for stars with $\tilde{\rm B}_{\odot} < 0^{\circ}$, 
and additionally includes the Pan-STARRS RC measurements by
\cite{S13}, shown as open pink rhombs; as well as the individual RRL
stars (small grey dots) and their mean distances in bins of
$\tilde{\Lambda}_{\odot}$ (filled lilac circles).

\begin{figure*}
  \includegraphics[width=0.47\textwidth]{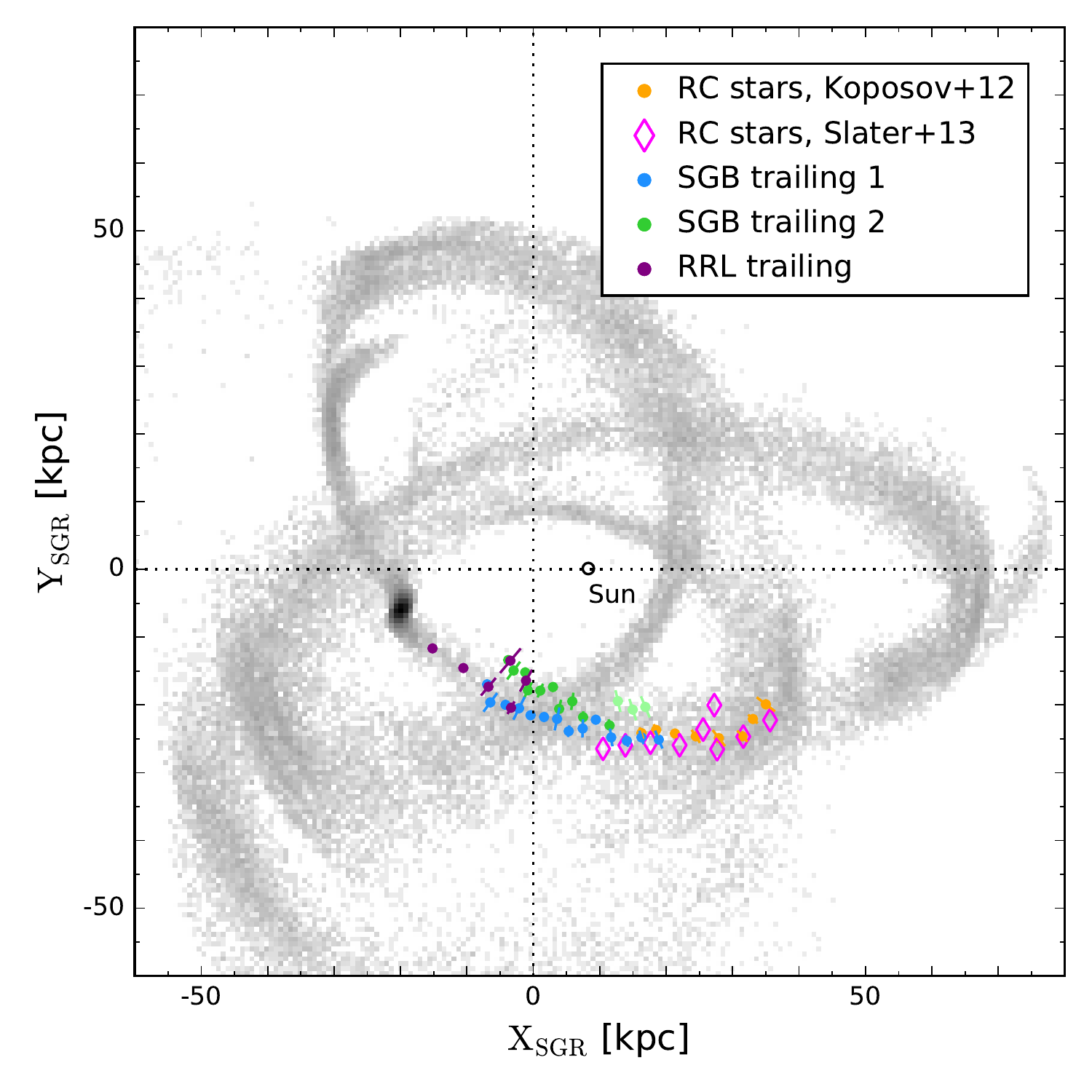}
  \includegraphics[width=0.47\textwidth]{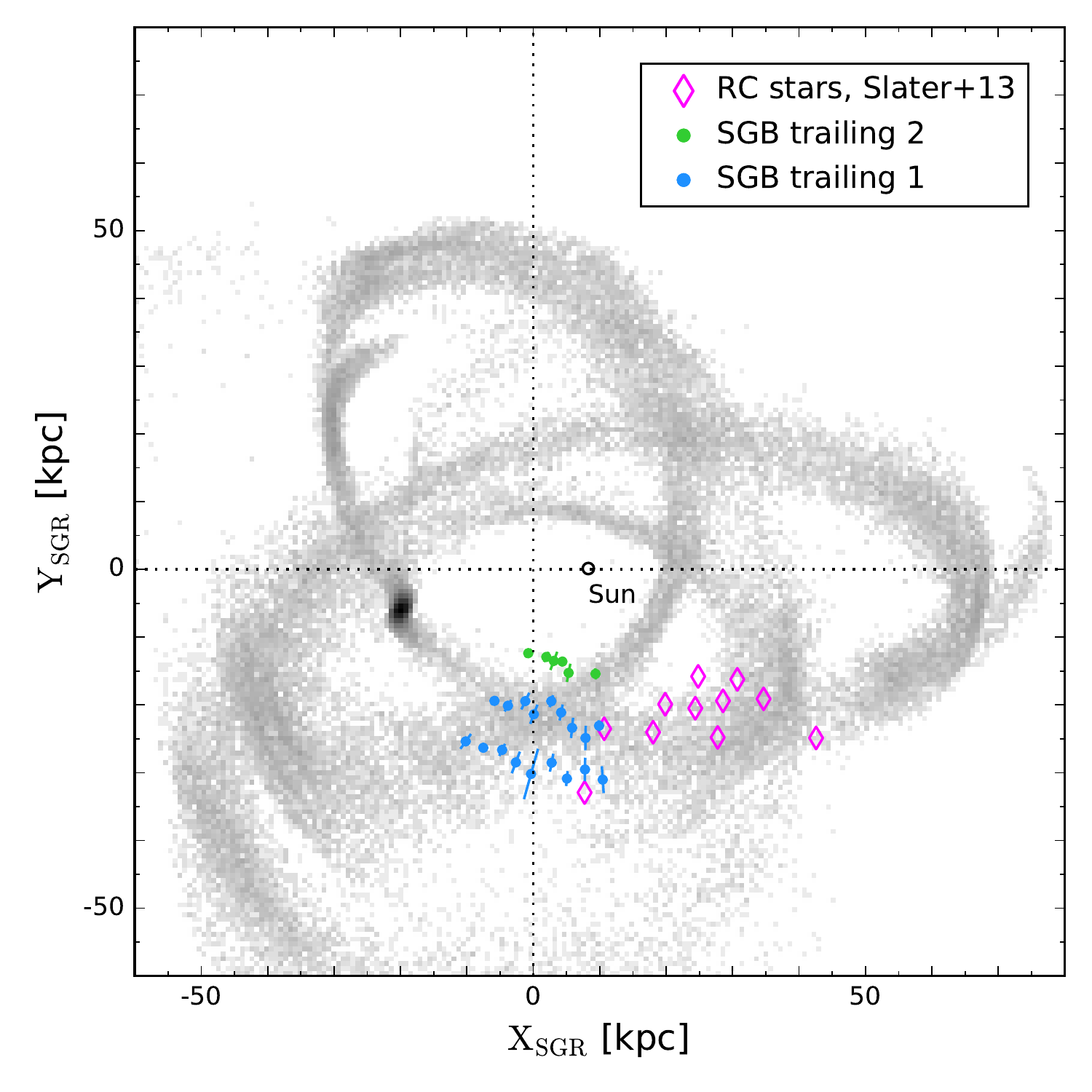}
  \caption{Centroids of the SGB detections, projected on the plane of the Sgr 
orbit, with the bright (faint) branch shown in the left (right) panel. The pole 
of the plane is at Galactocentric coordinates $l_{GC}$ = 275$^{\circ}$ and 
$b_{GC}$ = --14$^{\circ}$. The symbols are the same as in Figure~\ref{dist}, but 
the individual RRL stars from Sgr~1 are not shown here. The particles from the 
Sgr disruption model by \protect\cite{LM10} are shown as grey-scale density.}
    \label{xy}
\end{figure*}

The right panel of Figure~\ref{dist} gives the distances along the
faint trailing stream branch, i.e. that at $ 1^{\circ} < \tilde{\rm
  B}_{\odot} < 9^{\circ}$. The ``SGB trailing 1'' detections
correspond to the most prominent SGB peaks found in each
$\tilde{\Lambda}_{\odot}$ bin. Unlike the primary SGB signal found in
the Sgr bright arm (see the left panel of the Figure), this component
appears to split into two, almost parallel, branches separated by
$\sim$5 kpc. Curiously, there are hints of the two streams in the
previous measurements of \cite{S13}. However, that work lacked the
appropriate sky coverage to trace the stream over a long
$\tilde{\Lambda}_{\odot}$ range. As discussed in the previous Section,
at $\sim$15 kpc, there appear strong hints of another, albeit less
prominent sub-structure, dubbed ``SGB trailing 2'', which we believe is
the counterpart of the secondary SGB detection at similar distances in
the bright branch. Note that the multiple SGB detections along the
line of sight are corroborated by the BHB star counts as a function of
distance modulus. For example, the three substructures found in the
faint Sgr branch are also recovered using BHB stars (see the right panel of
Figure~\ref{bhbs}), where two prominent peaks are found at m-M $\sim$
16.8 and 17.3 mag (corresponding to the two trailing SGB 1), while a
less prominent peak is found at m-M $\sim$ 16.1 mag (SGB 2).
Unfortunately, the number of CRTS RRL stars at $\tilde{\rm B}_{\odot} >
0^{\circ}$ is not sufficient to trace the faint Sgr stream at this
stage.

Interestingly, the heliocentric distances reached by the most distant SGB 1 
debris in the faint Sgr branch are similar to the distances reported for the CPS 
by \cite{Newberg2009, Yam13}. While superficially these detections might appear 
related, as shown in Figure~\ref{bhbs}, the orbit of the CPS is not consistent 
with the track of these sub-structures \citep[see also Figure 8 from][]{S13}.

\subsection{In the Sgr orbital plane}

Figure~\ref{xy} displays the projection of the Sgr stream detections onto the 
Sgr orbital plane \citep[see][for more details]{B14}. Also shown as the 
underlying greyscale density is the distribution of the Sgr particles from the 
numerical model of \cite{LM10}. As is obvious from the left panel of the Figure, 
there is good agreement between the SGB 1 distance measurements within the 
bright branch of the trailing arm (shown in blue) and the simulation. The 
sequence of the SGB 1 detections and its continuation with RRL (lilac) point 
directly towards the Sgr dwarf. In comparison, the SGB 2 sub-structure does not 
have an obvious counterpart in the \cite{LM10} simulation. The two stream 
components ~--SGB 1 and SGB 2~-- start to diverge at $X_{\rm Sgr} \sim$ 0 kpc, 
and as indicated by the detections within the VST ATLAS, the SGB 2 structure 
does not seem to be going towards the Sgr remnant. Curiously, the two components 
diverge at the point where the young trailing debris cross the wrap of the 
leading tail. This, however, might be a pure coincidence given that the 
simulated leading tail appears to be moving almost perpendicular to the SGB 2.

Similarly, the right panel of Figure~\ref{xy} compares the orbital plane 
distribution of the simulated Sgr debris with the SGB detections of the faint 
branch of the trailing tail. The most prominent of the two SGB 1 detections in 
the faint arm, located at the heliocentric distance of $\sim$ 25 kpc, lies on 
top of the simulated trailing arm and points directly towards the Sgr dwarf 
nucleus. It connects with the previous detections made by \cite{S13}, extending 
the portion of the stream traced by an additional $\sim$30 kpc or 40 degrees. 
The other two SGB detections do not appear in the distribution of the simulated 
debris. However, again, around the location studied here, there appears to be a 
large spray of the leading debris.

 The projection of the Sgr stream detections onto the Sgr's
  orbital plane conceals the extension of the bright/faint streams
  below/above the plane in kpc. To clarify, the detections of the
  bright branch, confined to $-7^{\circ} < \tilde{\rm B}_{\odot} <
  0^{\circ}$, correspond to $Z_{\rm SGR} > -3.5$ and $-2.9$ kpc for
  SGB 1 and 2, respectively. SGB 1a, 1b and 2, in the faint branch,
  have $Z_{\rm SGR}$ up to 3.9, 5.0 and 2.5 kpc above the Sgr's plane
  (equivalent to $1^{\circ} < \tilde{\rm B}_{\odot} < 9^{\circ}$). In
  all the cases, the detections are confined to less than 5 kpc (in
  absolute value), which is a short distance considering the extension
  covered in the Sgr's plane coordinates ${\rm X_{\rm SGR}}$ and ${\rm
    Y_{\rm SGR}}$.

\subsection{Adding the third dimension}
\label{sec:3rd}

\begin{figure*}
  \centering
  \includegraphics[width=0.9\textwidth]{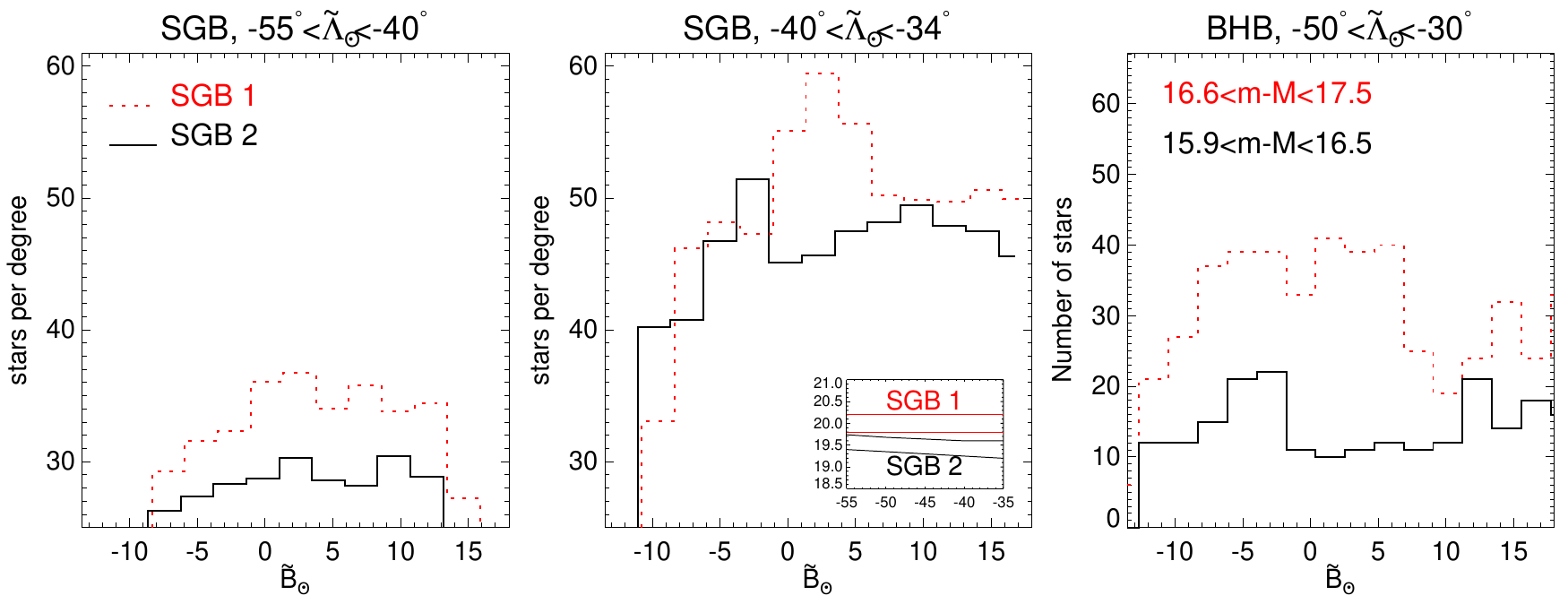}
  \caption{{\it Left and Middle}: Density distribution of the
    candidate SGB stars in the two line-of-sight components (SGB 1 and
    SGB 2) as a function of the stream latitude $\tilde{\rm
      B}_{\odot}$. The red dotted (black solid) histograms represent
    the SGB 1 (SGB 2) component. The SGB stars are selected to have
    0.52 $< (g-i) <$ 0.6. Additionally, a cut in the space of $i$
    magnitude vs. $\tilde{\Lambda}_{\odot}$ is applied and the exact
    selection boxes are shown in the inset of the middle panel. Note
    the different locations of the peaks of the two distributions in
    the middle panel, i.e. the portion of the stream closest to the
    Sgr remnant. {Right:} Density distribution of the candidate BHB
    stars. The main (secondary) structure corresponding to the SGB 1
    (SGB 2) component is shown with dotted red (solid black)
    line. Across the three panels, the solid black and the dotted red
    profiles show significant peaks around $\tilde{\rm
      B}_{\odot}=0^{\circ}$, thus lending support to our hypothesis
    that the structures we see along the line of sight are confined to
    the Sgr orbital plane, and therefore are likely connected to the
    accretion of the Sgr dwarf. Moreover, in the second and third
    panels, the secondary streams traced by the SGB and the BHB stars
    (black line) are off-set by a similar amount from zero,
    i.e. $\sim-5^{\circ}$, thus providing further evidence that the
    two stellar tracers are picking up the same structure.}
    \label{accross}
\end{figure*}

So far, we have limited our study to the two-dimensional slices of the
stream, i.e. we have only analyzed the stellar debris densities in the
plane of the distance and the stream longitude. This, of course, is a
necessity created by the complete overlap of the structures we have
discovered on the celestial plane. Figure~\ref{accross} presents an
attempt to discern whether the more distant overdensity (SGB 1) and
the closer one (SGB 2) behave differently as a function of the Sgr
latitude $\tilde{\rm B}_{\odot}$. For our experiment, we have chosen the
piece of the trailing arm where the number counts in the SGB 2 are at
the highest level, i.e. $-55^{\circ} <\tilde{\Lambda}_{\odot} <
-34^{\circ}$. Furthermore, we have split this section on the stream
into two portions shown in the left and the middle panels of the
Figure. The selection boundaries used to pick candidate stars
belonging to the SGB 1 and SGB 2 sub-structures are shown in the inset of
the middle panel of the Figure. In addition to the magnitude-longitude
cuts, all stars are also required to have colors consistent with the
SGB population, namely $0.52 < g-i < 0.6$.

According to Figure~\ref{accross}, there is a dramatic increase in the
number counts of stars in the SGB boxes from $\tilde{\Lambda}_{\odot}
= -55^{\circ}$ to $\tilde{\Lambda}_{\odot} = -34^{\circ}$, i.e. from
the left to the middle panel. Most of the density hike is associated with the
increase in the contaminating foreground population, but there appears
to be also an increase in the stream star counts. Note that while in
the right panel of Figure \ref{trace_sgb} the SGB 1 and SGB 2
overdensities have similar strength, in the histograms presented in
Figure~\ref{accross} the SGB 1 signal (shown in red) is clearly
stronger. This is because here the foreground counts were not
subtracted. The foreground contamination increases with apparent
magnitude, thus causing the SGB 1 signal to be elevated compared to
SGB 2. There exist obvious peaks in both the SGB 1 and the SGB 2
distributions, with locations close to
$\tilde{\rm B}_{\odot}\sim0^{\circ}$ ~-- i.e. in the vicinity of
the Sgr orbital plane~-- thus supporting the hypothesis that both
structures detected here used to belong to the Sgr dwarf proper.

Perhaps, the most interesting feature of the Figure is the difference in
the latitudinal behavior of the two substructures at $-40^{\circ}
<\tilde{\Lambda}_{\odot} < -34^{\circ}$ displayed in the middle
panel. The more distant SGB 1 appears to have a fairly narrow peak at
$\tilde{\rm B}_{\odot}=2.5^{\circ}$. This can be compared to the black
histogram representing the SGB 2 counts, where the still narrow peak is
shifted by several degrees towards negative $\tilde{\rm B}_{\odot}$. One
possible, albeit tentative, explanation of the difference in the
latitudinal profiles is that SGB 1 connects directly to the progenitor
while SGB 2 might not. Note that there is another, broader peak in
the distribution of the SGB 2 candidate stars, at
$\tilde{\rm B}_{\odot}\sim10^{\circ}$. Further away from the remnant,
i.e. at $-55^{\circ} <\tilde{\Lambda}_{\odot} < -40^{\circ}$, no
discernible difference can be seen in the stellar density profiles of
the two sub-structures.

Finally, the right panel of Figure~\ref{accross} presents the
$\tilde{\rm B}_{\odot}$ density profile of structures traced by BHBs
with different distance moduli. Following the above notation, red
dotted line corresponds to the main (SGB 1, or $16.6 <m-M<17.5$)
component and solid black to the closer (SGB 2, or $15.9 < m-M<16.5$)
one. As expected, the red line traces a broad bump, centered around
$\tilde{\rm B}_{\odot}=0^{\circ}$. The black histogram displays a peak
slightly off $\tilde{\rm B}_{\odot}=0^{\circ}$ location, at around
$\tilde{\rm B}_{\odot}=-3^{\circ}$, similarly to the SGB 2 profile
shown in the middle panel of the Figure. Given that there is a very
good match between the peak positions and the widths of the density
distributions of the bright SGB stars (SGB 2) and the bright BHBs, we
conclude that it is likely that both sets of tracers correspond to the
same halo structure. Furthermore, taking into account the proximity of
the density peaks to $\tilde{\rm B}_{\odot}=0^{\circ}$, it is very
likely that this structure is related to the Sgr disruption.

\subsection{Metallicity of the detected sub-structures}

RRL stars from the catalogue of \cite{T15} have photometrically
derived metallicities available, thus giving us an opportunity for a
straightforward comparison of the chemical properties of the two
stream components. As usual, the metallicities of the individual RRL
stars were derived using the light curve shape information. The
RRL-based [Fe/H] distribution of a nearby piece of the trailing tail
was studied before by \citet{W09}, who reported a mean metallicity of
[Fe/H]=--1.43 with a dispersion of 0.3 dex. For the portion of the tail
covered by the CRTS, we measure average metallicity of [Fe/H]$=
-1.47$, with a dispersion of $\sigma$ = 0.36 dex. Clearly, our
estimate of the Sgr 1 mean metallicity is very similar to the
abundance of the Sgr tail covered by the SDSS Stripe 82, as derived by
\cite{W09}. However, it remains uncertain how much of this agreement
is due to a genuine similarity of the two portions of the stream
studied and how much of it may be due to the crudeness of the RRL
photometric [Fe/H] scale.

An indication that the agreement may not be fortuitous is provided by the
spectroscopic follow-up study of the Sgr trailing RRL stars detected
by the SEKBO survey as reported by \citet{P09}. Here, a more
metal-poor mean abundance of [Fe/H] = --1.79$\pm$ 0.08 was found with
a dispersion of 0.38 dex. Therefore, spectroscopic metallicities for
RRL stars belonging to Sgr\,1 are needed to confirm if they are indeed
more metal-rich than dictated by the findings of \cite{P09}. Note,
however that our mean metallicities are in very good agreement with
those found for a nearby portion of the stream in the most recent
spectroscopic study of \citet{Gibbons2016}. They report a bimodal 
MDF, in which the more prominent metal-poor component has the mean at
--1.33 and a dispersion of 0.27 dex.

Additionally, we have compared the mean metal abundances of the two RRL 
overdensities detected in the first two $\tilde{\Lambda}_{\odot}$ bins. It turns 
out that the two overdensities have the same mean metal abundances: at 
$\tilde{\Lambda}_{\odot} = -43.5^{\circ}$, [Fe/H] = --1.53 (with dispersion 
$\sigma$=0.2 dex) for the 9 RRL at 19.2 kpc, while the 11 RRL at $\sim$24 kpc 
have [Fe/H] $= -1.52$ ($\sigma$ = 0.3 dex); at $\tilde{\Lambda}_{\odot} = 
-31.5^{\circ}$, 9 RRL at 18 kpc have [Fe/H] $= -1.51$ ($\sigma$ = 0.4 dex), 
while the 12 RRL at 22 kpc have [Fe/H]$= -1.49$ ($\sigma$ = 0.3 dex). It appears 
therefore that both line-of-sight components detected in the bright Sgr trailing 
branch are consistent with the metallicities expected for the Sgr stars. 
Clearly, more information, such as radial velocities of the RRL stars in each of 
the peaks, would be required to confirm the components as coherent and related 
structures. Such information will be provided by our team in a forthcoming 
study (Duffau et al., in preparation).

\subsection{Conclusions}

In this work, we have examined a large portion~-- approximately
$65^{\circ}$~-- of the Sgr trailing stream available in the imaging
data from the VST ATLAS survey. Most of the area~-- at least
$40^{\circ}$~-- covered in our analysis has not been studied before
using photometric data of such depth. This section of the stream is of
particular interest as it is situated in the proximity of the Sgr
remnant, i.e. only $40^{\circ}$ from the center of the dwarf. Taking
advantage of the depth of the VST ATLAS photometry, we chose to use
the SGB population as the main halo sub-structure tracer.

Curiously, at many locations along the Sgr stream studied here, at
least two peaks of SGB stars are detected along the line of sight. At
its highest, the separation between the peaks is $\sim$0.5 mag (or
$\sim$5 kpc at the distance of the stream). However, importantly, we
detect a significant variation in the peak separation as a function of
the stream longitude $\tilde{\Lambda}_{\odot}$. Therefore, we believe
that the secondary SGB detection is not due to a complex co-distant
stellar population mix in the stream, but rather reveals the presence
of (at least) two distinct sub-structures projected within the Sgr
orbital plane. This discovery is reminiscent of the detection of the
so-called Branch C behind the Sgr leading tail presented in
\citet{B06}.

We have compared, where possible, the absolute and the relative SGB distance 
measurements with the previously published values such as those by \cite{K12} 
and \cite{S13} and found very good agreement. Furthermore, we have confirmed the 
authenticity of the SGB line-of-sight detections with other tracers, such as RR 
Lyrae and BHB stars, and were reassured by an excellent match. The two 
line-of-sight components are most visible in the bright branch of the Sgr 
trailing stream, i.e. at $\tilde{\rm B}_{\odot}<0^{\circ}$. Nonetheless, we also 
find strong evidence for a similar line-of-sight splitting in the faint branch, 
i.e. at $\tilde{\rm B}_{\odot}>0^{\circ}$. 

The projection of the VST ATLAS detections onto the Sgr orbital plane reinforces 
the conclusion that the two sub-structures are indeed part of the Sgr debris 
distribution. When viewed in this perspective, the more distant of the two 
overdensities, the SGB 1, appears to be connecting to the Sgr remnant. The two 
sub-structures can not be easily separated at large distances from Sgr; however, 
the presence of an additional stream component is betrayed by the increased 
width of the line-of-sight distribution of the debris. On approach to the dwarf, 
however, at $\tilde{\Lambda}_{\odot}\sim-60^{\circ}$, the SGB 2 forks out, and, 
following a distinct distance gradient, appears to undershoot the Sgr dwarf. In 
comparison to the simulation of \citet{LM10}, the bifurcation happens around the 
location where the young trailing debris runs into an old wrap of the leading 
arm. This poses a question whether the SGB 2 could actually be one of the old 
wraps of the Sgr stream. Note, however, that there is no obvious
counterpart of the SGB 2 sub-structure in the above mentioned simulation; 
therefore, at the moment, this scenario seems somewhat unlikely.

With the new detections presented here, the current picture of the Sgr
tidal tails appears incredibly complex. The trailing arm is ripped
apart on the plane of the sky, but additionally shows a bifurcation
along the line of sight. If these, previously unseen portions of the
Sgr debris were indeed the older wraps of the stream, they would
provide yet another confirmation that the Sgr tails are longer than
previously thought, and are likely originating from a bigger parent
galaxy. Additionally, these new wraps ought to provide a good leverage
in constraining the properties of the gravitational potential of the
Galaxy. Alternatively, the messy twisting of the tidal tails could be
a sign of the complex structure of the progenitor. For example,
\citet{Jorge2010} show that line-of-sight stream splitting is
possible if the Sgr dwarf was a disc galaxy. \citet{Simon2016}
confirms that even in a completely spherical host potential, the discy
dwarf disruption will produce sprays of tidal debris in distinct
orbital planes. Clearly, to establish the nature of the multiple
sub-structures within the Sgr tails, it would help to focus on the part
of the stream so far virtually unstudied, namely the section closest
to the progenitor.

\section*{Acknowledgements}

This publication makes use of data products from the AAVSO Photometric
All Sky Survey (APASS), funded by the Robert Martin Ayers Sciences
Fund and the National Science Foundation.

The research leading to these results has received funding from the
European Research Council under the European Union's Seventh Framework
Programme (FP/2007-2013)/ERC Grant Agreement no. 308024.

This project is supported by CONICYT's PCI program through grant
DPI20140066.  Additional support is provided by the Ministry for the
Economy, Development, and Tourism's Iniciativa Cient\'ifica Milenio
through grant IC\,120009, awarded to the Millennium Institute of
Astrophysics; by Proyecto Fondecyt Regular \#1141141; and by Proyecto
Basal PFB-06/2007. C.N. acknowledges support from CONICYT-PCHA grant
Doctorado Nacional 2015-21151643.




\bibliographystyle{mnras}
 \bibliography{biblio}





%
%


\bsp	
\label{lastpage}
\end{document}